\documentclass[a4paper,11pt]{article}
\usepackage{jcappub} 

\usepackage{hyperref}
\hypersetup{
  colorlinks=true,        
  linkcolor=blue,         
  citecolor=cyan,         
}

\usepackage{graphicx}
\usepackage{dcolumn}
\usepackage{bm}
\usepackage{color}
\usepackage{enumitem}
\usepackage{amsmath}
\usepackage{amssymb}
\usepackage{cleveref}
\usepackage{orcidlink}
\usepackage{graphicx} 
\usepackage{subfigure}
\usepackage{soul,xcolor}
\setstcolor{red}

\crefname{section}{Sec.}{Secs.}      
\Crefname{section}{Sec.}{Secs.}      
\crefname{Figure}{Fig.}{Figs.}      
\Crefname{Figure}{Fig.}{Figs.}      

\arxivnumber{2508.21519}
\title{Lensing Stability and Scattering Phenomena in Anisotropic Black Hole Spacetimes in Plasma}

\author[a,b]{Jonibek Khasanov
\orcidlink{0009-0005-7875-1327}}
\affiliation[a]{Institute of Fundamental and Applied Research, National Research University TIIAME, Kori Niyoziy 39, Tashkent 100000, Uzbekistan}
\affiliation[b]{National Pedagogical University of Uzbekistan named after Nizami, Bunyodkor 27, Tashkent, Uzbekistan}

\author[c,d]{Mirzabek Alloqulov
\orcidlink{0000-0001-5337-7117}}
\affiliation[c]{School of Physics, Harbin Institute of Technology, Harbin 150001, China}
\affiliation[d]{Tashkent State Technical University, 100095 Tashkent, Uzbekistan}

\author[e]{Pankaj Sheoran
\orcidlink{0000-0001-8283-8744}}
\affiliation[e]{Department of Physics, School of Advanced Sciences, Vellore Institute of Technology, Tiruvalam Rd, Katpadi, Vellore, Tamil Nadu 632014, India}

\author[a,f,g,h]{Sanjar Shaymatov
\orcidlink{0000-0002-5229-7657}}
\affiliation[f]{University of Tashkent for Applied Sciences, Str. Gavhar 1, Tashkent 100149, Uzbekistan}
\affiliation[g]{Samarkand State University, University Avenue 15, Samarkand, 140104, Uzbekistan}
\affiliation[h]{Western Caspian University, Baku AZ1001, Azerbaijan}

 \author[i]{Hemwati Nandan
\orcidlink{0000-0002-1183-4727}
}
\affiliation[i]{Department of Physics, Hemvati Nandan Bahuguna Garhwal Central University, Srinagar Garhwal, Uttarakhand 246174, India}
\emailAdd{khasanovjonibek775@gmail.com}
\emailAdd{malloqulov@gmail.com}
\emailAdd{pankaj.sheoran@vit.ac.in}
\emailAdd{sanjar@astrin.uz}
\emailAdd{hnandan@associates.iucaa.in}

\abstract{We investigate the physical and observational features of static, spherically symmetric black hole spacetimes surrounded by anisotropic fluid and embedded in a plasma environment. Motivated by recent advances in black hole imaging and precision measurements in strong gravity, we explore light propagation, wave dynamics, and observational signatures in such geometries. We begin by analyzing the background spacetime and matter content, examining the horizon structure and verifying the energy conditions associated with the anisotropic fluid. We then study photon trajectories in both vacuum and plasma environments, deriving the equations of motion and computing deflection angles and image magnifications under weak gravitational lensing. Both uniform and power-law plasma profiles are considered to model realistic astrophysical settings. In the wave optics regime, we first analyze the linear stability of the spacetime under axial (odd-parity) perturbations using Chandrasekhar’s method, and then investigate scalar field scattering by solving the wave equation in the curved background with plasma. Using Born and WKB approximations, we compute the differential scattering cross-sections and examine how anisotropy and plasma affect interference features. Finally, we perform parameter estimation using Markov Chain Monte Carlo methods to constrain the black hole mass and anisotropic fluid parameters, utilizing EHT and GRAVITY data for Sgr A* and EHT-only data for M87*. These results present a unified theoretical framework that links anisotropic matter effects with lensing and scattering observables providing a firm basis for future comparisons with high-resolution astrophysical data sets in diverse contexts.}  
\compress
\begin{document}
\maketitle

\section{Introduction}

Black holes, once considered purely theoretical constructs, are now firmly embedded in observational astrophysics. The landmark imaging of the supermassive black holes at the centers of M87 and the Milky Way (Sgr A*) by the Event Horizon Telescope (EHT)~\cite{EventHorizonTelescope:2019dse, EventHorizonTelescope:2022wkp} and the precision tracking of stellar orbits around Sgr A* by the GRAVITY collaboration~\cite{Gillessen:2008qv, Hees:2017aal, GRAVITY:2021xju,GRAVITY:2024tth} have opened new avenues for testing gravitational theories in the strong-field regime. { Crucially, these observations rely on interferometric measurements in the visibility domain, where the signatures of higher-order images and photon rings provide powerful probes of spacetime geometry \cite{Vincent:2022fwj,Aratore:2021usi,Feleppa:2025ejh}. The analysis of visibility amplitudes and phases from VLBI arrays represents what is actually measured observationally, making this an essential consideration for connecting theoretical predictions with experimental data.} Complementing these electromagnetic observations, the direct detection of gravitational waves by LIGO and Virgo~\cite{LIGOScientific:2016aoc,KAGRA:2021vkt} has revolutionized our ability to probe the dynamical behavior of black holes, including their mergers and ringdown phases. { Additional high-impact observational methods include X-ray reflection spectroscopy \cite{Bambi:2016sac,Bambi:2020jpe}, which probes the inner accretion flow and strong gravity regime through analysis of the iron K$\alpha$ line and Compton hump, and techniques based on the orbital angular momentum (OAM) of light \cite{Tamburini:2021lyi,Tamburini:2021jok,Feleppa:2025pgg}, which provide novel ways to characterize spacetime curvature and matter distributions through the twisting of light wavefronts.} These multi-messenger signals provide not only direct evidence for the existence of event horizons but also invaluable insights into the geometry, stability, and near-horizon physics of compact objects \cite{Konoplya:2018yrp,Chen:2021fxr}. As observational techniques continue to advance, increasingly precise measurements of shadow sizes, lensing signatures \cite{Virbhadra:1999nm,Bozza:2010xqn,LIGOScientific:2021izm}, quasi-normal modes \cite{Konoplya:2011qq,Berti:2009kk}, and wave scattering demand theoretical models that extend beyond standard vacuum solutions like Schwarzschild or Kerr \cite{PhysRevD.16.937,PhysRevD.18.1798,PhysRevD.31.1869,Glampedakis:2001cx}, and that incorporate realistic matter distributions and environmental effects such as surrounding plasma.

A natural extension entails black holes surrounded by anisotropic fluids—matter fields whose radial and tangential pressures differ \cite{1995JMP....36..340D,PhysRevD.62.104002,PhysRevD.96.064053}. Unlike ideal perfect fluids, anisotropic fluids can arise in a variety of physically motivated settings \cite{Kumar:2017tdw,PhysRevD.104.104039}. These include exotic field configurations in early-universe cosmology \cite{Watanabe:2009ct,Ackerman:2007nb}, models of dark energy  \cite{Yang:2018ubt} and dark matter halos \cite{Bharadwaj:2003iw,Al-Badawi25CPC}, quantum gravitational corrections \cite{Fernandes:2023vux,Gambini:2020qhx}, and effective descriptions of astrophysical accretion flows \cite{refId0,Kurmanov:2021uqv}. Anisotropy is also a natural feature in the presence of strong magnetic fields \cite{Foucart:2015cws}, phase transitions \cite{Caprini:2010xv}, or particle creation processes \cite{Lima:2007kk}. The presence of such matter can influence the black hole's horizon structure \cite{Kiselev:2002dx,Shaymatov21d,deMToledo:2018tjq,Shaymatov18a,Chabab:2020ejk,Shaymatov21pdu,Shaymatov22a}, modify photon spheres \cite{C:2024cnk}, alter deflection angles of light \cite{Kumar:2017tdw,Mustafa22CPC}, and impact wave dynamics \cite{Jusufi:2019ltj,Al-Badawi25CTP_DM,Alloqulov25EPJC}. Importantly, these effects may leave observable imprints in the form of lensing distortions, shadow deformations, and interference patterns in wave scattering, which are detectable through electromagnetic \cite{EventHorizonTelescope:2019dse} and gravitational wave observations \cite{Jusufi:2019ltj}.

Motivated by all these, this work focuses on a static, spherically symmetric black hole solution sourced by an anisotropic fluid characterized by two parameters: \( K \), which quantifies the strength of the anisotropic matter, and \( \omega_2 \), the equation-of-state parameter controlling its radial decay \cite{Cho19ChPhC..43b5101C}. The resulting metric introduces a deviation from Schwarzschild geometry while preserving spherical symmetry and asymptotic flatness (or de Sitter behavior depending on the sign of \( K \)). This framework allows us to interpolate between well-known physical regimes: radiation-like matter with \( \omega_2 = 1/3 \), pressureless dust with \( \omega_2 = 0 \), and dark energy-like behavior for \( \omega_2 < 0 \). Such flexibility enables a unified analysis of various astrophysical scenarios within a single geometric setup.

{Gravitational lensing emerges} as one of the most promising observational signatures in this context. When light from a distant source passes near a compact object, its trajectory is deflected due to spacetime curvature. This effect is not only a classic test of general relativity but also a sensitive probe of the geometry near compact objects \cite{Bozza:2010xqn}. In the presence of anisotropic fluids, the deflection angle and lensing magnification are modified in characteristic ways \cite{Azreg-Ainou:2017obt}. Moreover, realistic lensing scenarios often involve the presence of plasma \cite{Bisnovatyi-Kogan:2010flt,Bisnovatyi-Kogan:2017kii,Atamurotov:2021cgh,2013Ap&SS.346..513M,Bisnovatyi-Kogan:2010flt}, which introduces frequency-dependent refraction and can alter both the deflection angle and brightness profile of lensed images. {Recent analytical treatments of light propagation in dispersive media have significantly advanced our understanding of plasma lensing effects \cite{Perlick:2017fio,Perlick:2023znh,Feleppa:2024vdk}, providing rigorous frameworks for studying frequency-dependent ray tracing and intensity modifications in astrophysical plasmas. These developments are particularly relevant for interpreting radio observations from instruments like the EHT, where plasma effects cannot be neglected.} Incorporating plasma—both uniform and non-uniform—is therefore essential for making meaningful comparisons with radio observations such as those from the EHT.

Beyond classical deflection, wave phenomena provide an additional window into black hole physics. Scattering of scalar \cite{PhysRevD.79.064022,10.1063/1.522949,Glampedakis:2001cx}, electromagnetic \cite{PhysRevD.7.2807,PhysRevD.12.933,Crispino:2009xt}, or gravitational waves \cite{Dolan:2008kf,OLeary:2008myb,Vishveshwara:1970zz} in curved spacetime reveals information about quasi-normal modes, stability, and near-horizon geometry \cite{Kokkotas:1999bd,Konoplya:2019xmn}. The presence of anisotropic matter and plasma alters the effective potential experienced by these waves, modifying the scattering cross-section and interference pattern. Partial wave analysis, especially using Born \cite{Batic:2011aa} and WKB \cite{1985ApJ...291L..33S,PhysRevD.35.3632,Simone:1991wn} approximations, enables us to compute these observables and identify regimes where anisotropy or plasma effects become significant. These calculations are both theoretically insightful and relevant for interpreting signals from pulsars near galactic centres as well as gravitational wave echoes from black hole mergers.

Finally, the increasing availability of high-resolution observational data offers an unprecedented opportunity to test and constrain modified black hole metrics. Using Bayesian parameter estimation techniques such as Markov Chain Monte Carlo (MCMC), one can infer the most likely values of the model parameters \( M \), \( K \), and \( \omega_2 \) consistent with current data. In this work, we use EHT observations of M87* and a combination of EHT and GRAVITY data for Sgr A* to fit our anisotropic black hole model. This approach allows us to test for potential deviations from the Schwarzschild case and assess whether anisotropic matter components could be present in the vicinity of real astrophysical black holes.

Taken together, our study presents a comprehensive analysis of an anisotropic fluid black hole model, bridging geometric structure, stability, lensing observables, wave scattering, and data-driven parameter inference. By connecting theoretical predictions with measurable quantities, we aim to provide a framework for probing new physics in the strong gravity regime, constrained by current and future observations.

This study is structured as follows: In \cref{Sec:Spacetime}, we introduce the black hole spacetime sourced by an anisotropic fluid, analyzing its geometric structure, horizon formation, and matter properties. We verify the energy conditions and investigate linear stability under axial gravitational perturbations to ensure the dynamical viability of the spacetime. \cref{sec:EOM} presents the equations governing photon motion, derived using the Hamiltonian formalism, and incorporates the effects of both uniform and non-uniform plasma profiles that are relevant in realistic astrophysical environments. In \cref{sec:w_lensing}, we study weak gravitational lensing by computing the deflection angles of light rays influenced by the combined effects of anisotropic fluid and plasma. This sets the stage for \cref{Sec:magnification}, where we analyze the magnification of lensed images, highlighting how plasma and anisotropic matter alter the observed brightness of background sources. In \cref{Sec:WaveDynamics}, we turn to wave dynamics, examining the scattering of massless scalar fields in the anisotropic black hole background. We compute the differential scattering cross-section using both the Born and WKB approximations for various plasma environments, revealing how interference patterns encode information about the spacetime and surrounding medium. Finally, in \cref{sec:param_estimation_anisotropic}, we perform parameter estimation \cite{Foreman-Mackey_2013} using observational data from the Event Horizon Telescope (EHT) and GRAVITY collaboration. Specifically, we constrain the black hole parameters \( M \), \( K \), and \( \omega_2 \) by fitting lensing observables to EHT-only data for M87* and combined EHT+GRAVITY data for Sgr~A*. This integrated approach provides a framework to assess the observational signatures of anisotropic matter near black holes and test deviations from standard Schwarzschild geometry. In the end, we conclude our study and highlight the key outcomes of our analysis in \cref{sec:conlusion}.

Throughout this study, we employ the natural system of units with \( G = c = \hbar = 1 \), unless stated otherwise. Dimensional quantities such as mass, length, and frequency are expressed in geometrized units. We also use the metric signature \((- + + +)\), consistent with standard general relativity conventions. The black hole mass parameter $M=1$ is assumed in all numerical plots for the sake of simplicity. In the section on parameter estimation (Section~VII), we restore appropriate physical units to facilitate direct comparison with observational data from the Event Horizon Telescope (EHT) and GRAVITY collaboration. This approach allows us to connect theoretical predictions with real astrophysical measurements, assess whether the signatures of anisotropic matter are compatible with existing observations, and provide a theoretical reference for future efforts to identify or constrain such matter near black holes using high-resolution astrophysical data.

\section{Anisotropic Black Hole Spacetime and Matter Conditions}
\label{Sec:Spacetime}

We start by examining the geometry of a black hole surrounded by anisotropic fluid and study its horizon structure and energy conditions. These foundational results will inform our subsequent exploration of photon motion and black hole shadows in the presence of plasma.

\subsection{Metric Structure and Horizon Analysis}

{Here, we briefly review a black hole spacetime with an anisotropic fluid environment developed by the authors of Ref.~\cite{Cho19ChPhC..43b5101C}, addressing the solution of the Einstein field equation to find exact solutions in the presence of an anisotropic fluid. To this end, the energy-momentum tensor for an anisotropic fluid with the corresponding metric for the general case is written as follows:
\begin{eqnarray}
    T_{\mu \nu}=(\rho+p_2)u_{\mu}u_{\nu}+(p_1-p_2)x_{\mu}x_{\nu}+p_2g_{\mu\nu}\, ,
\end{eqnarray}
with the energy density $\rho$ and its timelike four-velocity $u^\mu$ and a spacelike unit vector $x^\mu$ that is orthogonal to the four-velocity $u^\mu$ and angular directions, including the equation of state with the following relation $p_i=\omega_i \rho$, where $w_i< -1$ refers to the phantom energy, $w_i< -1/3$ to the dark energy, $w_i=0$ to the dust, and $w_i=1/3$ to the radiation. In order to find analytic solutions of the Einstein field equation, the exactly solvable case was considered by setting $w_1=-1$, leading to the static and spherically symmetric spacetime in the following form (see details, e.g.~\cite{Cho19ChPhC..43b5101C})}
\begin{equation}
    ds^2 = -f(r)\,dt^2 + \frac{dr^2}{f(r)} + r^2 \left( d\theta^2 + \sin^2{\theta}\,d\phi^2 \right),
\end{equation}
with the metric function defined as
\begin{equation}\label{f(r)}
    f(r) = 1 - \frac{2M}{r} - \frac{K}{r^{2\omega_2}}.
\end{equation}
Here, \( M \) denotes the ADM mass of the black hole, \( K \) is a parameter linked to the energy density of the anisotropic fluid, and \( \omega_2 \) is the equation-of-state parameter characterizing the fluid. The parameters \( K \) and \( \omega_2 \) crucially influence the causal structure and asymptotic behavior of the geometry.

{Certain values of \( \omega_2 \) correspond to familiar matter types:}
\begin{description}
    \item[Case I: \( \omega_2 = 0 \)] Dust-like matter, yielding an effective cosmological constant term;
    \item[Case II: \( \omega_2 = \frac{1}{3} \)] Radiation, producing a potential that decays slowly with \( r \);
    \item[Case III: \( \omega_2 = -\frac{1}{2} \)] Dark energy-like behavior, introducing a repulsive term that grows with \( r \).
\end{description}

\begin{figure*}[!htb]
    \centering
    \includegraphics[width=\textwidth]{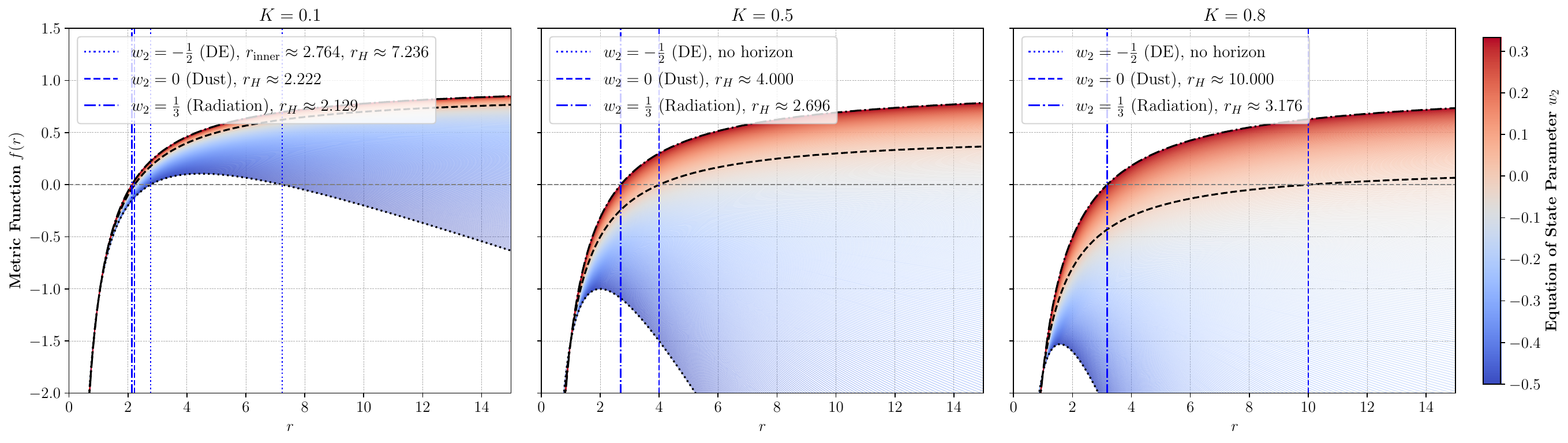}
    \caption{Behavior of the metric function \( f(r) \) for a black hole surrounded by anisotropic fluid, shown for three different values of the fluid parameter \( K = 0.1, 0.5, 0.8 \) (from left to right). The curves correspond to a range of equation-of-state parameters \( \omega_2 \in [-\frac{1}{2}, \frac{1}{3}] \). Special cases \( \omega_2 = -\frac{1}{2} \) (dark energy), \( \omega_2 = 0 \) (dust), and \( \omega_2 = \frac{1}{3} \) (radiation) are shown in black, with vertical blue lines marking the corresponding event horizons \( r_H \).}
    \label{fig:anisotropic_metric_K_scan}
\end{figure*}

To illustrate the behavior of the metric function \( f(r) \), Figure~\ref{fig:anisotropic_metric_K_scan} shows its variation for three representative values of \( K = 0.1, 0.5, 0.8 \), across a range of \( \omega_2 \in [-\frac{1}{2}, \frac{1}{3}] \). Special cases (\( \omega_2 = -\frac{1}{2}, 0, \frac{1}{3} \)) corresponding to dark energy, dust, and radiation are highlighted in black, while vertical blue lines indicate the associated event horizon radii \( r_H \). The ADM mass is fixed at \( M = 1 \).

To determine the location of the event horizons, we solve \( f(r) = 0 \), yielding
\[
1 - \frac{2M}{r} - \frac{K}{r^{2\omega_2}} = 0.
\]
Multiplying through by \( r^{2\omega_2} \), we obtain the general form
\[
r^{2\omega_2} - 2M r^{2\omega_2 - 1} - K = 0.
\]
The above equation becomes algebraically tractable for special choices of \( \omega_2 \):

\vspace{0.5cm}
\noindent\textbf{Case I: \( \omega_2 = 0 \) (Dust)}
\[
f(r) = 1 - \frac{2M}{r} - K \quad \Rightarrow \quad r_H = \frac{2M}{1 - K}, \quad \text{valid for } K < 1.
\]

\begin{itemize}
    \item One horizon for \( K < 1 \),
    \item No horizon (naked singularity) for \( K \geq 1 \).
\end{itemize}

\vspace{0.5cm}
\noindent\textbf{Case II: \( \omega_2 = \frac{1}{3} \) (Radiation)}
\[
f(r) = 1 - \frac{2M}{r} - \frac{K}{r^{2/3}}, \quad \text{let } x = r^{1/3} \Rightarrow x^3 - Kx - 2M = 0.
\]
This is a depressed cubic whose discriminant determines the number of real roots:
\[
\Delta = -4K^3 - 108M^2.
\]
For clarity, the possible horizon configurations for different signs of $K$ (with $M>0$) are summarized in table \ref{tab:horizon_structure}. As seen from the table, a positive $K$ always yields a single horizon, whereas $K<0$ allows the discriminant  to become positive, leading to three real roots and hence multiple horizons.

\begin{table}[htb]
\centering
\footnotesize
\renewcommand{\arraystretch}{1.1}
\resizebox{\columnwidth}{!}{%
\begin{tabular}{|c|c|c|}
\hline
\textbf{Case} & \textbf{Discriminant \( \Delta \)} & \textbf{Horizon Structure} \\
\hline
\( M > 0,\, K > 0 \) & \( \Delta < 0 \) & One real root \( \Rightarrow \) One horizon \\
\hline
\( M > 0,\, K < 0 \) & \( \Delta > 0 \) possible & Three real roots \( \Rightarrow \) Multiple horizons \\
\hline
\end{tabular}%
}
\caption{Horizon structure for different signs of \( K \) with \( M > 0 \) in the radiation case.}
\label{tab:horizon_structure}
\end{table}

\vspace{0.5cm}
\noindent\textbf{Case III: \( \omega_2 = -\frac{1}{2} \) (Dark Energy-like)}
\[
f(r) = 1 - \frac{2M}{r} - K r \quad \Rightarrow \quad K r^2 - r + 2M = 0,
\]
with roots
\[
r = \frac{1 \pm \sqrt{1 - 8KM}}{2K}.
\]
\begin{itemize}
    \item Two horizons if \( 8KM < 1 \),
    \item One degenerate horizon if \( 8KM = 1 \),
    \item No horizon if \( 8KM > 1 \).
\end{itemize}

\begin{figure}[htbp]
    \centering
    \includegraphics[scale=0.5]{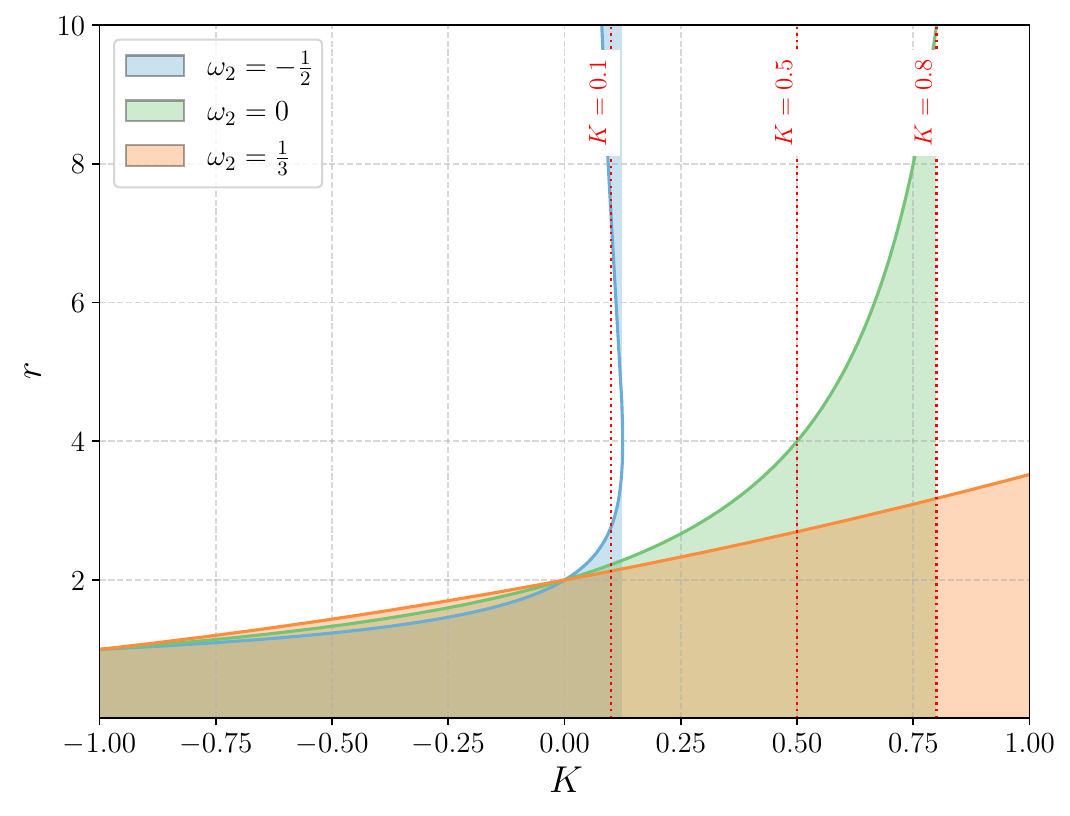}
    \caption{
        Parameter space {between parameter $K$ and $r$ illustrating} the shaded regions where the metric function \( f(r) < 0 \) for an anisotropic black hole spacetime, corresponding to three values of the equation-of-state parameter: \( \omega_2 = -\frac{1}{2} \), \( \omega_2 = 0 \), and \( \omega_2 = \frac{1}{3} \). 
        The shaded regions are plotted only when a horizon exists, i.e., when \( f(r) = 0 \) has at least one positive root. 
         The vertical red dotted lines mark specific values of \( K = 0.1,\, 0.5,\, 0.8 \), which are used to guide interpretation.
        The solid contours in matching colors indicate the location of the horizon where \( f(r) = 0 \).
    }
    \label{fig:anisotropic_BH_shaded}
\end{figure}

Figure~\ref{fig:anisotropic_BH_shaded} visualizes the parameter space in the \( (K, r) \) plane, showing the shaded regions where horizons exist (\( f(r) = 0 \) has real, positive roots). Solid contours trace the horizon locations, and vertical red dotted lines at \( K = 0.1,\, 0.5,\, 0.8 \) match the values used in Figure~\ref{fig:anisotropic_metric_K_scan}.

\subsection{Energy Conditions of the Anisotropic Fluid}

To evaluate the physical plausibility of the anisotropic matter content, we examine the standard energy conditions based on the energy-momentum tensor:
\begin{equation}
    T^{\mu}_{~\nu} = \text{diag}(-\rho, p_r, p_t, p_t),
\end{equation}
where $\rho$ is the energy density, $p_r$ is the radial pressure, and $p_t$ is the tangential pressure.

Using Einstein's equations, the energy-momentum components for the metric function \( f(r) \) yield:
\begin{align}
    8\pi \rho(r) &= \frac{1}{r^2} \left(1 - f - r f' \right), \\
    8\pi p_r(r) &= \frac{1}{r^2} \left(f - 1 + r f' \right), \\
    8\pi p_t(r) &= \frac{1}{4r} \left(2 f' + r f'' \right),
\end{align}
with derivatives
\begin{align}
    f'(r) &= \frac{2M}{r^2} + \frac{2\omega_2 K}{r^{2\omega_2 + 1}}, \\
    f''(r) &= -\frac{4M}{r^3} - \frac{2\omega_2(2\omega_2 + 1)K}{r^{2\omega_2 + 2}}.
\end{align}

The standard energy conditions are:
\begin{itemize}
    \item \textbf{Null Energy Condition (NEC)}: $\rho + p_r \geq 0$, $\rho + p_t \geq 0$;
    \item \textbf{Weak Energy Condition (WEC)}: $\rho \geq 0$, $\rho + p_r \geq 0$, $\rho + p_t \geq 0$;
    \item \textbf{Strong Energy Condition (SEC)}: $\rho + p_r + 2p_t \geq 0$;
    \item \textbf{Dominant Energy Condition (DEC)}: $\rho \geq |p_r|$, $\rho \geq |p_t|$.
\end{itemize}

Representative results:
\begin{description}
    \item[Case I] $\omega_2 = 0$ (dust), the metric reduces to a Schwarzschild–(A)dS type. The energy conditions are satisfied for $K < 0$ (de Sitter-like).
    \item[Case II] $\omega_2 = \frac{1}{3}$ (radiation), the $K$-term decays slowly with $r$, modifying intermediate radial profiles.
    \item[Case III] $\omega_2 = -\frac{1}{2}$ (dark energy-like), the $K$-term increases with $r$. For suitable signs and magnitudes of $K$, the NEC and WEC are preserved, though the SEC can be violated—mimicking dark energy behavior.
\end{description}

For physically reasonable ranges of the fluid parameters \( K \) and \( \omega_2 \), the anisotropic matter content satisfies the null and weak energy conditions across the relevant spacetime domain. This lends physical credibility to the black hole solutions under consideration, ensuring that the underlying geometry is sourced by viable matter configurations. With the spacetime structure and its matter content established, we now proceed to examine the propagation of light in this background, with particular focus on gravitational lensing effects arising from the interplay between the black hole geometry, the anisotropic fluid, and surrounding plasma.


\section{Equations of motion}\label{sec:EOM}

We employ the Hamilton–Jacobi formalism to investigate photon trajectories in the spacetime of a black hole surrounded by anisotropic fluid. The Hamiltonian describing null geodesics in the presence of plasma can be expressed as~\cite{Benavides_Gallego_2019}:
\begin{equation}
\mathcal H(x^\alpha, p_\alpha)=\frac{1}{2}\left[ g^{\alpha \beta} p_\alpha p_\beta - (n^2-1)( p_\beta u^\beta )^2 \right]\ ,
\label{generalHamiltonian}
\end{equation}
where $x^{\alpha}$ is the spacetime coordinates, $u^{\beta}$ and $p_{\alpha}$ refer to the four-velocity and momentum of the photon, respectively. $n=\omega/k$ is the refractive index with wave number $k$. 

{The equations of motion for light rays in the presence of plasma require careful treatment due to frequency-dependent effects. Recent analytical developments in understanding light propagation in dispersive media provide important foundations for our approach \cite{Perlick:2017fio,Perlick:2023znh,Feleppa:2024vdk}. The Hamiltonian formalism we employ here is particularly suited for handling the refractive effects introduced by both uniform and non-uniform plasma distributions around black holes.}

One can write it in the following form~\cite{alfandari2020approximationdoubletravellingsalesman}
\begin{eqnarray}
n^2&=&1- \frac{\omega_{\text{p}}^2}{\omega^2}\, ,
\label{eq:n1}
\end{eqnarray}
with $\omega_p$ and $\omega$ are the plasma and photon frequencies, respectively. We can define the photon frequency using the $\omega^2=(p_{\beta}u^{\beta})^2$ {following}:
\begin{equation}
\omega(r)=\frac{\omega_0}{\sqrt{f(r)}}\ ,\qquad  \omega_0=\text{const}\, .
\end{equation}
with $f(r) \to 1$ is satisfied as $r \to \infty$ and $\omega(\infty)=\omega_0=-p_t$ ~\cite{Perlick15a}.

\begin{figure}[htbp]
    \includegraphics[scale=0.285]{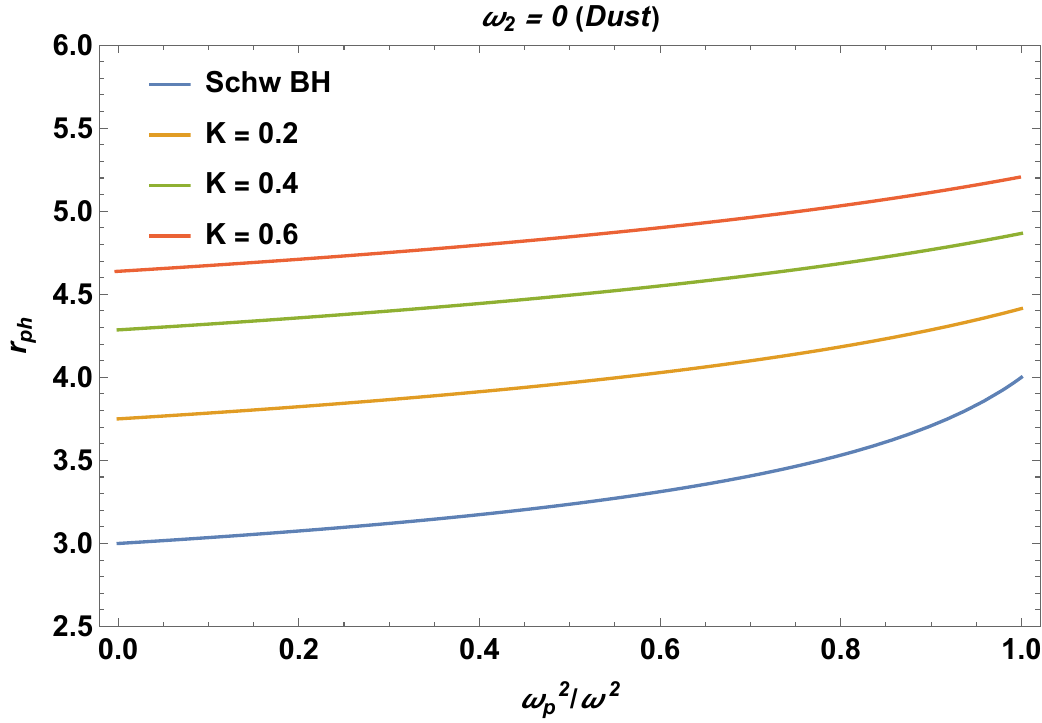}
     \includegraphics[scale=0.285]{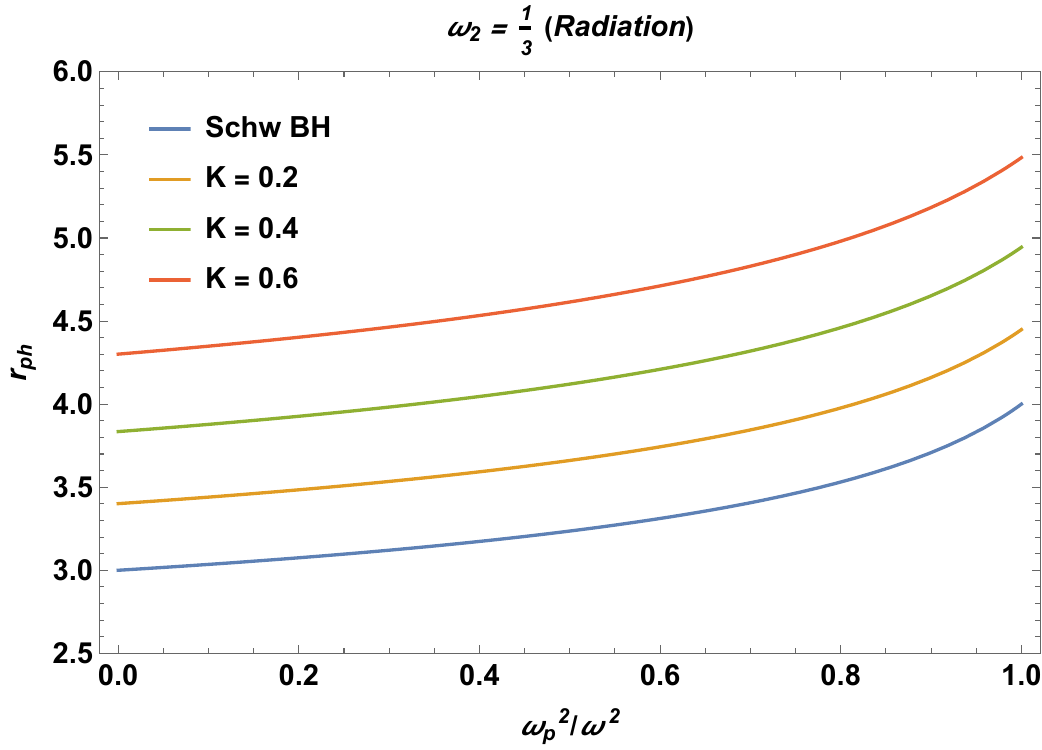}
     \includegraphics[scale=0.285]{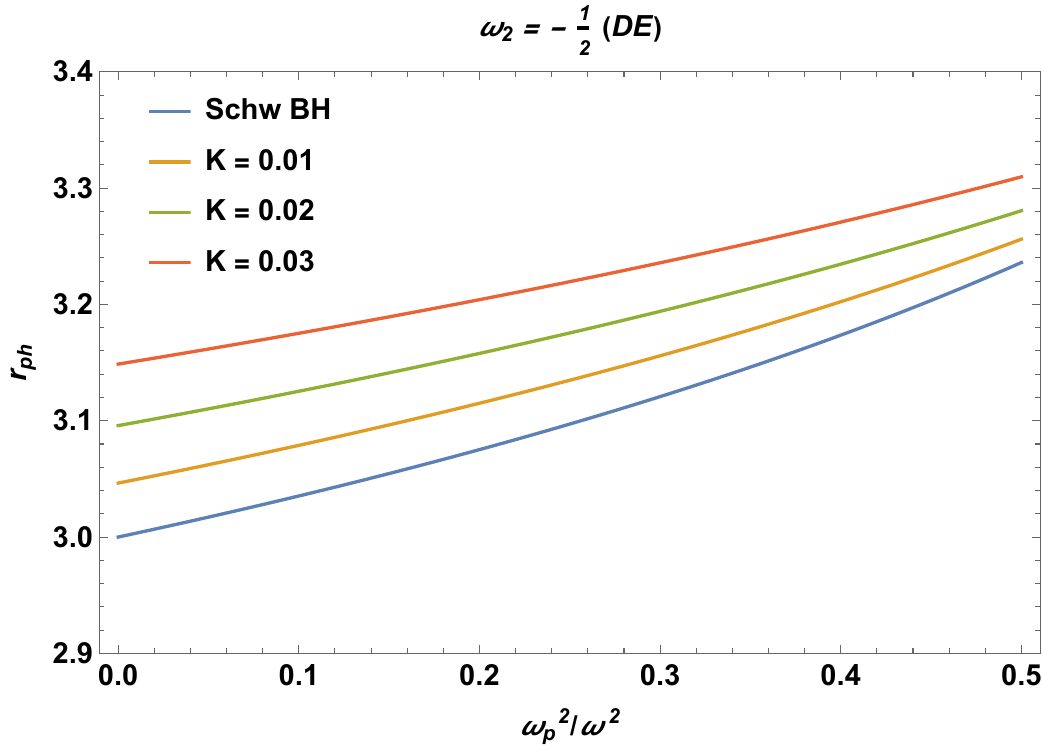}
    \caption{The photon sphere radius as a function of plasma frequency for different values of the $K$ parameter. {Note that all three (dust, radiation and DE) cases have been considered.}}
    \label{fig:ph}
\end{figure}

One can rewrite the Eq.(\ref{generalHamiltonian}) by considering the above equations as~\cite{Synge:1960b,Rog:2015a}
\begin{equation}
\mathcal{H}=\frac{1}{2}\Big[g^{\alpha\beta}p_{\alpha}p_{\beta}+\omega^2_{\text{p}}]\, . \label{eq:hamiltonnon}
\end{equation}
The light ray equations in the equatorial plane ($\theta=\pi/2$) can be written using the above Hamiltonian as
\begin{eqnarray} 
\dot t\equiv\frac{dt}{d\lambda}&=& \frac{ {-p_t}}{f(r)}\, , \label{eq:t} \\
\dot r\equiv\frac{dr}{d\lambda}&=&p_r f(r)\, , \label{eq:r} \\
\dot\phi\equiv\frac{d \phi}{d\lambda}&=&  \frac{p_{\phi}}{r^2}. \label{eq:varphi}
\end{eqnarray}
The orbit equation can be derived by using Eqs.~(\ref{eq:r}) and (\ref{eq:varphi}) as
\begin{equation}
\frac{dr}{d\phi}=\frac{g^{rr}p_r}{g^{\phi\phi}p_{\phi}}\, .    \label{trajectory}
\end{equation}
The above equation takes the following form for the light geodesics ($\mathcal{H}=0$)
\begin{equation}
 \frac{dr}{d\phi}=\sqrt{\frac{g^{rr}}{g^{\phi\phi}}}\sqrt{\gamma^2(r)\frac{\omega^2_0}{p_\phi^2}-1}\, ,
\end{equation}
where
\begin{equation}\label{eq:hrnew}
    \gamma^2(r)\equiv-\frac{g^{tt}}{g^{\phi\phi}}-\frac{\omega^2_p}{g^{\phi\phi}\omega^2_0}\ . 
\end{equation}
It is worth noting that the photon comes from infinity and reaches a minimum at a radius $r_{ph}$, and then goes to infinity again. From the mathematical point of view, this radius corresponds to a turning point of the $\gamma^2(r)$ function. Therefore, one can find the photon radii from the following relationship:
\begin{equation}
\frac{d(\gamma^2(r))}{dr}\bigg|_{r=r_{\text{ph}}}=0\, . \label{eq:con}    
\end{equation}
We perform a numerical calculation for the photon sphere and show its dependence on plasma frequency for different values of the spacetime parameters in Fig.~\ref{fig:ph}. {We can see from this figure that the values of the photon sphere radii increase with the increase of the $K$ parameter for the three cases, which are the dust, radiation and dark energy-like cases.}

\textit{Black hole shadow in plasma:} 
\begin{figure}[htbp]
    \includegraphics[scale=0.285]{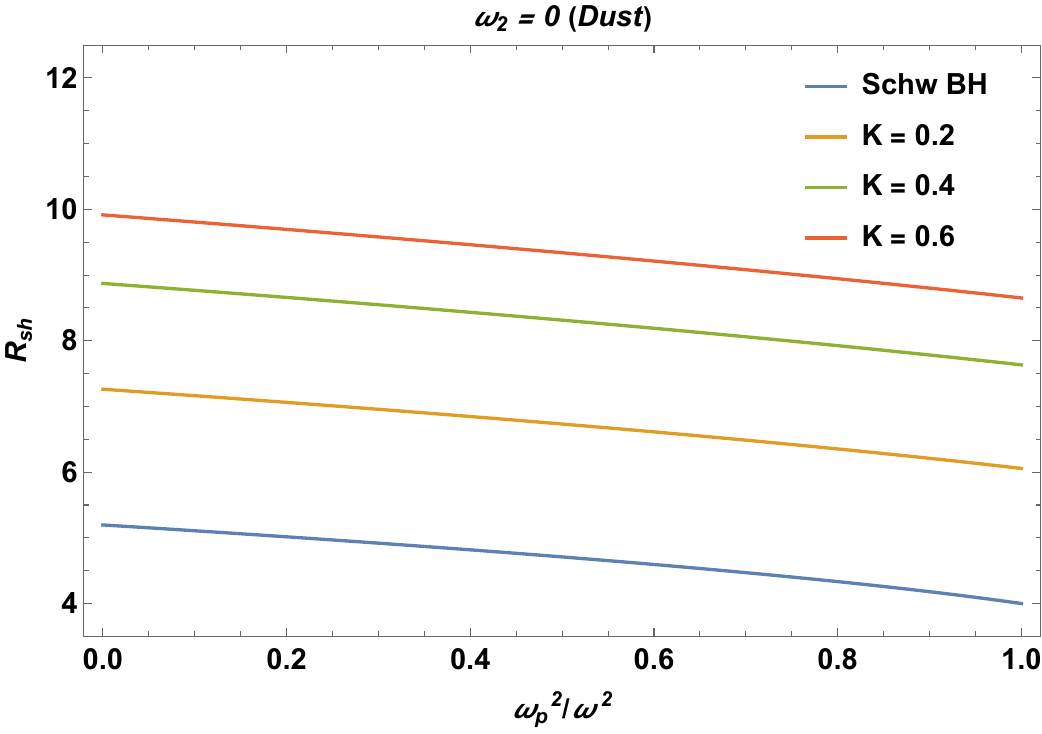}
    \includegraphics[scale=0.285]{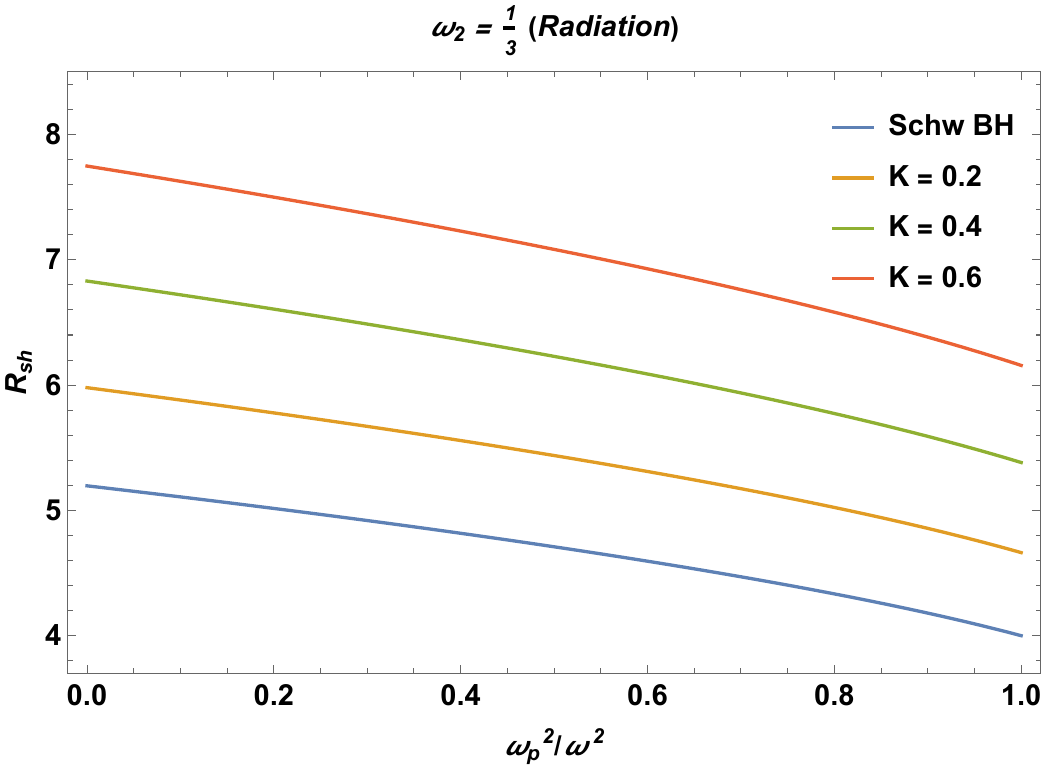}
    \includegraphics[scale=0.285]{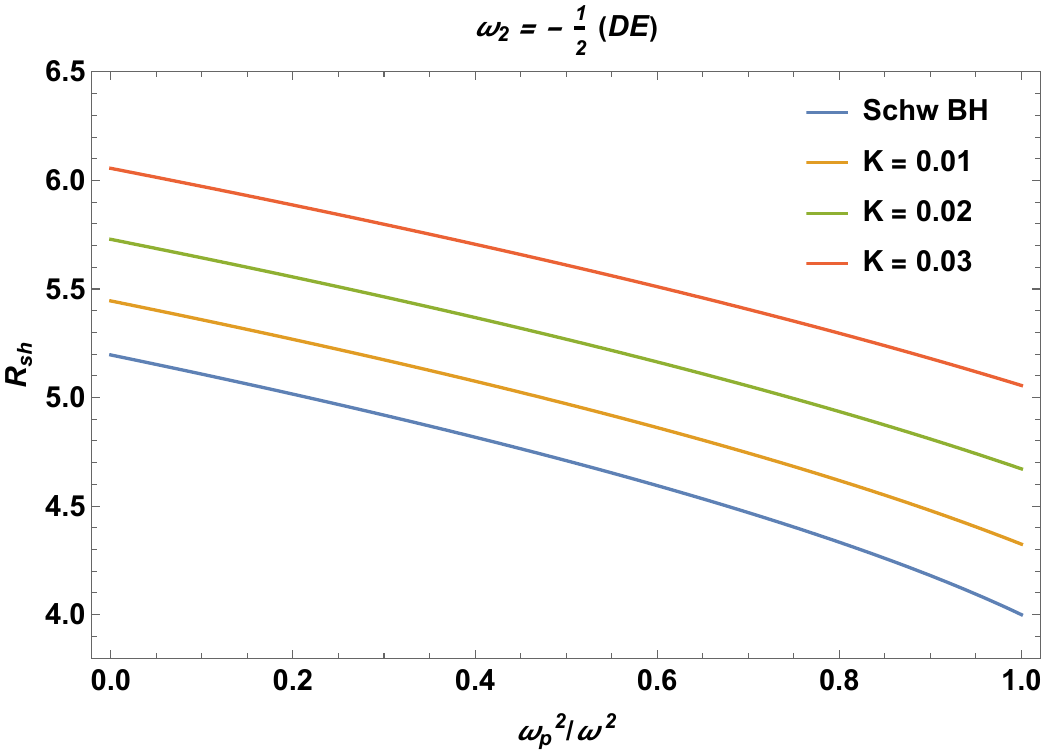}
    \caption{The radius of the black hole shadow with the plasma frequency for different values of the $K$ parameter. {Note that all three (dust, radiation and DE) cases have been considered.}}
    \label{fig:shadow}
\end{figure}
\begin{figure*}
\includegraphics[scale=0.54]{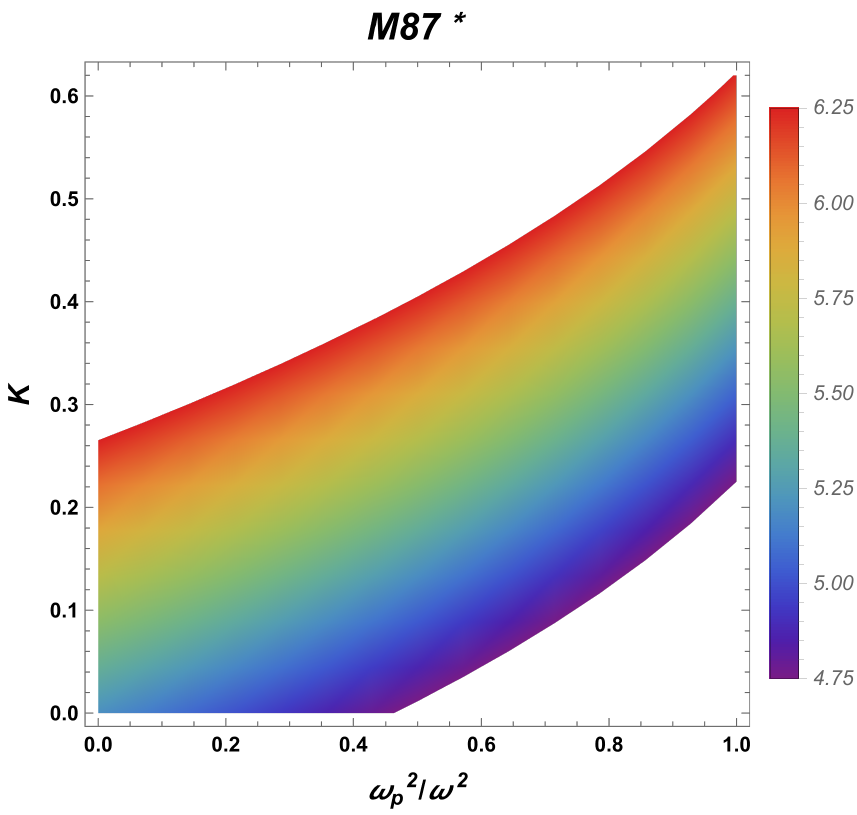}
\includegraphics[scale=0.54]{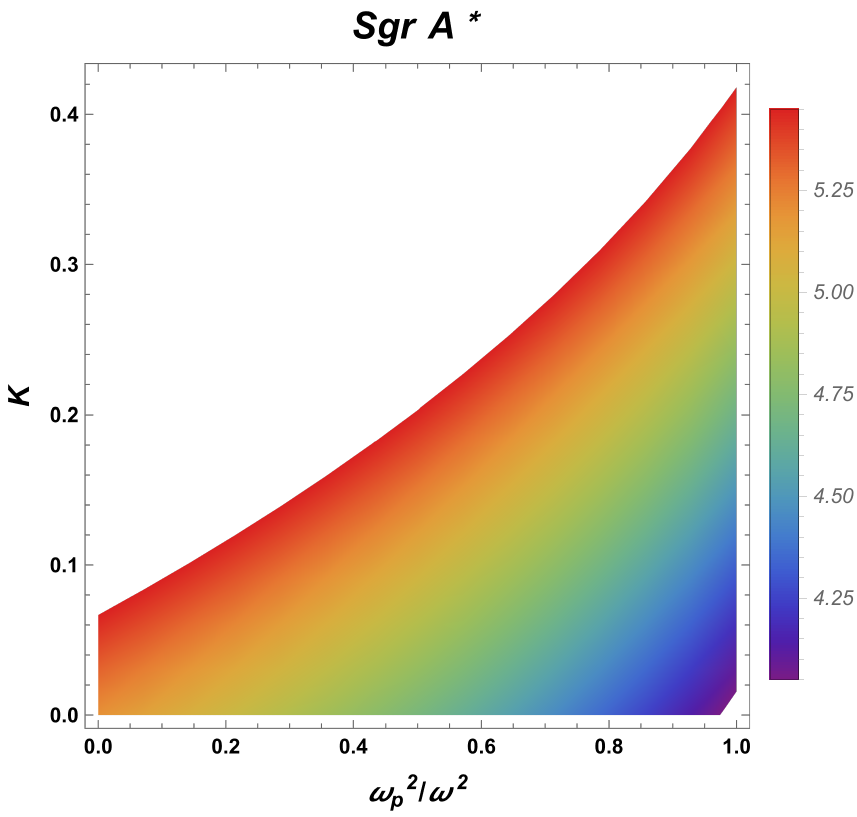}
\caption{\textcolor{black}{{The constrained values of the parameter
$K$  as a function of plasma frequency for the supermassive black holes M87* and Sgr A*.}}}
\label{fig:constraint}
\end{figure*}
In this part, we explore the shadow of the black hole with an anisotropic fluid. Firstly, we can write the angular radius of the black hole shadow $\alpha_{sh}$ in the following form~\cite{Perlick15a,Konoplya2019}
\begin{eqnarray}
\label{eq:shadow nonrotating1}
\sin^2 \alpha_{\text{sh}}&=&\frac{\gamma^2(r_{\text{ph}})}{\gamma^2(r_{\text{o}})},\nonumber \\&=&\frac{r^2_{\text{ph}}\left[\frac{1}{f(r_{\text{ph}})}-\frac{\omega^2_p(r_{\text{ph}})}{\omega^2_0}\right]}{r^2_{\text{o}}\left[\frac{1}{f(r_{\text{o}})}-\frac{\omega^2_p(r_{\text{o}})}{\omega^2_0}\right]},
\end{eqnarray}
where $r_{ph}$ and $r_{o}$ are the locations of the photon sphere and observer, respectively. One can write the radius of the black hole shadow as follows by considering the assumption that the observer is located far from the black hole~\cite{Konoplya2019}
\begin{eqnarray}
R_{\text{sh}}&\simeq& r_{\text{o}} \sin \alpha_{\text{sh}},\\
 &=&\sqrt{r^2_{\text{ph}}\left[\frac{1}{f(r_{\text{ph}})}-\frac{\omega^2_p(r_{\text{ph}})}{\omega^2_0}\right]},  \nonumber
\end{eqnarray}
Using the above equation, we can explore numerically the black hole shadow. Fig.~\ref{fig:shadow} illustrates the radius of the black hole shadow as a function of the plasma frequency for different values of the spacetime parameters. {It can be seen from this figure that the values of the black hole shadow radii increase with the increase of the $K$ parameter for the three cases, which are the dust, radiation and dark energy-like cases.} What is more, we make an assumption that M87* and Sgr A* are static and spherically symmetric, even though the EHT collaboration does not support this.  One can investigate the limits of the $K$ parameter using the provided results by the EHT collaboration. It is worth noting that the $K$ parameter and plasma frequency were chosen for constraint. 
\begin{table}[htbp]
\centering
\caption{Observational Data for M87* and Sgr A* (VLTI)~\cite{EventHorizonTelescope:2019dse}.}
\label{tab:data}
\resizebox{\textwidth}{!}{
\begin{tabular}{|l|c|c|}
\hline
\textbf{Parameter}         & \textbf{M87*} & \textbf{Sgr A*} \\ \hline
Angular Diameter ($\theta$) & $42 \pm 3 \, \mu\text{as}$ & $48.7 \pm 7 \, \mu\text{as}$ \\ \hline
Distance ($D$)             & $16.8 \pm 0.8 \, \text{Mpc}$ & ${(82.77 \pm 0.33) \times 10^2 \, \text{pc}}$ \\ \hline
Mass ($M$)                 & $(6.5 \pm 0.7) \times 10^9 \, M_{\odot}$ & $(4.297 \pm 0.013) \times 10^6 \, M_{\odot}$ \\ \hline
\end{tabular}
}
\end{table}
The EHT results, including the angular diameter, the distance from Earth and the black hole mass for M87* and Sgr A*, are given in Table~\ref{tab:data}.  Using these results, we can calculate the diameter of the shadow caused by the compact object per unit mass as follows~\cite{Bambi_2019}:
\begin{equation}
    d_{sh}=\frac{D \theta}{M}.
\end{equation}
Finally, one can write the values of the diameter of the black hole shadow $d^{M87*}_{sh}=(11 \pm 1.5)M$ for M87* and $d^{Sgr*}_{sh}=(9.5 \pm 1.4)M$ for Sgr A*. Using the above information, we demonstrate the limits of the $K$ parameter and plasma frequency by ``color map" in Fig.~\ref{fig:constraint}. 
 \label{Sec:lensing}
\begin{figure}[!htb]
 \begin{center}
      \includegraphics[scale=0.4]{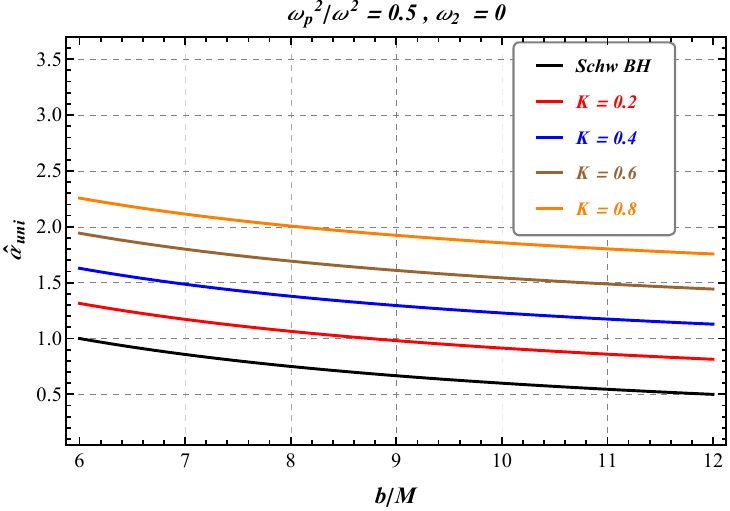}
   \includegraphics[scale=0.4]{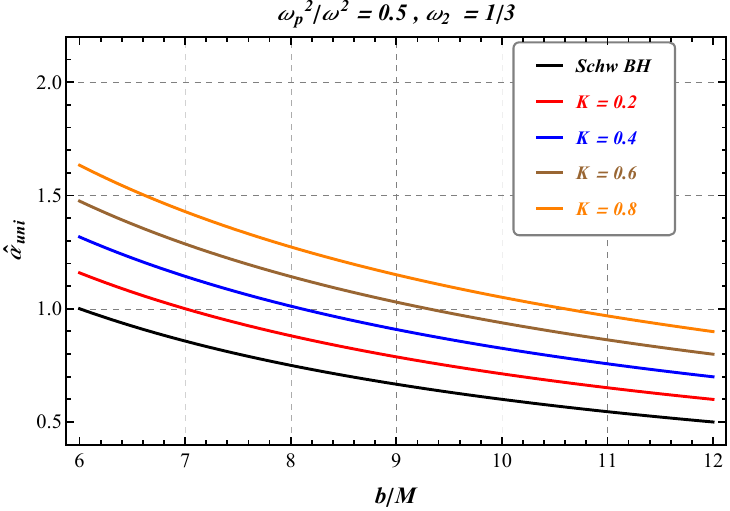}
   \includegraphics[scale=0.4]{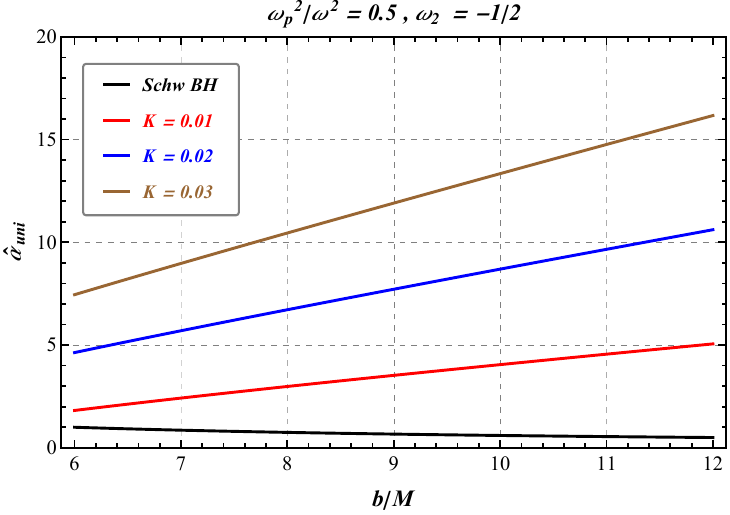}
  \end{center}
\caption{\label{fig:6} 
Variation of deflection angle $\hat{\alpha}_{\text{uni}}$ as a function of the impact parameter $b$ for different values of the parameter $K$ in a uniform plasma medium $\omega_{p}^2/\omega^2$. The left panel shows the case $w_{2}=0$, the middle panel corresponds to $w_{2}=1/3$, and the right panel presents the case $w_{2}=-1/2$.}
\end{figure}
\begin{figure}[!htb]
 \begin{center}
    \includegraphics[scale=0.4]{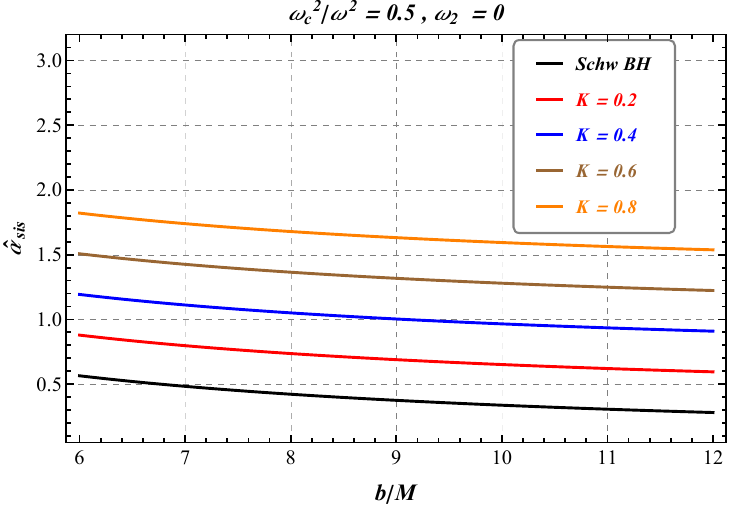}
   \includegraphics[scale=0.4]{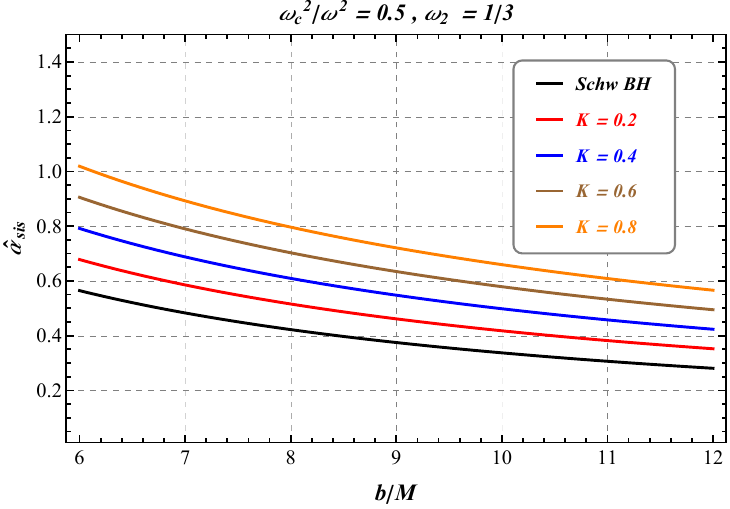}
   \includegraphics[scale=0.4]{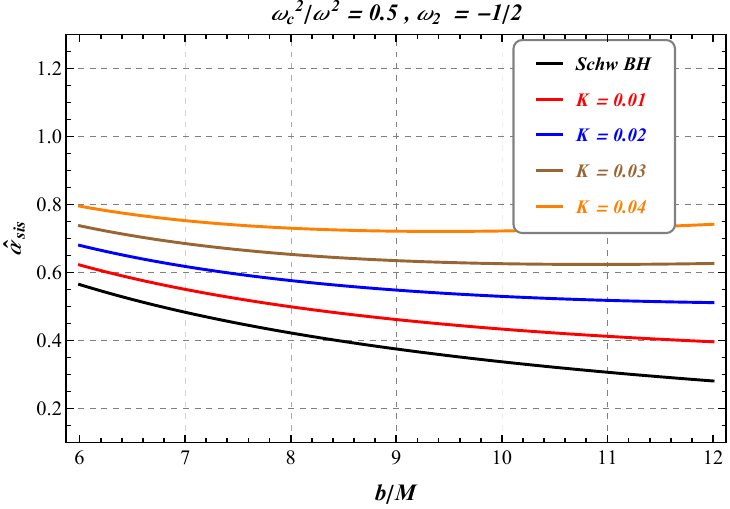}
  \end{center}
\caption{\label{fig:7} Variation of deflection angle $\hat{\alpha}_{\text{uni}}$ as a function of the impact parameter $b$ for different values of the parameter $K$ in a non-uniform plasma medium $\omega_{c}^2/\omega^2$. The left panel shows the case $w_{2}=0$, the middle panel corresponds to $w_{2}=1/3$, and the right panel presents the case $w_{2}=-1/2$.}
\end{figure}

\section{Weak gravitational lensing for a black hole with anisotropic fluid 
}
\label{sec:w_lensing}

{
While our analysis focuses on image-plane observables, it is crucial to recognize that actual VLBI observations measure interferometric visibilities in the Fourier domain. The lensing signatures we compute here would manifest as specific features in the visibility amplitudes and phases \cite{Vincent:2022fwj,Aratore:2021usi}. The modifications to deflection angles due to anisotropic matter could potentially be detected through detailed analysis of higher-order image signatures in VLBI data.
}

Here, we examine lensing effects in the weak form around an anisotropic black hole. To that, one needs to utilize a required expression describing the weak-field approximation (see, e.g.,~\cite{Bisnovatyi-Kogan:2010flt,2021PDU....3200798B})
\begin{equation}
    g_{\alpha \beta}=\eta_{\alpha \beta}+h_{\alpha \beta}\, ,
\end{equation}
with $\eta_{\alpha\beta}$ and $h_{\alpha\beta}$ representing a gravitational potential for the Minkowski spacetime and gravity field, respectively. With this in view and following~\cite{Bisnovatyi-Kogan:2010flt}, the following relation between these two potentials is written as follows: 
\begin{eqnarray}
 &&   \eta_{\alpha \beta}=diag(-1,1,1,1)\ , \nonumber\\
 &&   h_{\alpha \beta} \ll 1, \hspace{0.5cm} h_{\alpha \beta} \rightarrow 0 \hspace{0.5cm} under\hspace{0.2cm}  x^{\alpha}\rightarrow \infty \ ,\nonumber\\
 &&     g^{\alpha \beta}=\eta^{\alpha \beta}-h^{\alpha \beta}, \hspace{0,5cm} h^{\alpha \beta}=h_{\alpha \beta}\, .
\end{eqnarray}
Taking this relation into consideration, one is available to define the deflection angle of light moving around the black hole with an anisotropic fluid as  
\begin{eqnarray}
    \hat{\alpha }_{\text{b}}&=&\frac{1}{2}\int_{-\infty}^{\infty}\frac{b}{r}\left(\frac{dh_{33}}{dr}+\frac{1}{1-\omega^2_e/ \omega^2}\frac{dh_{00}}{dr}\right.\nonumber\\&-&\left.\frac{K_e}{\omega^2-\omega^2_e}\frac{dN}{dr} \right)dz\, , 
\end{eqnarray}
where we have defined $\omega_{e}$ and $\omega$ as the plasma and photon frequency, respectively. In the weak field regime, the spacetime metric expanded in the Taylor series yields as 
\begin{eqnarray}
  ds^2=ds^2_0 &+&\left(\frac{2 M}{r}+Kr^{-2w_2}\right)dt^2 \nonumber\\&+&(\frac{2 M}{r}+Kr^{-2w_2} )\frac {z^2}{r^2} dr^2\, , 
\end{eqnarray}
where $ds^2_0=-dt^2+dr^2+r^2(d\theta^2+\sin^2\theta d\phi^2)$. 
The components of the perturbation gravity field $h_{\alpha \beta}$ in Cartesian coordinates are written as follows:
\begin{eqnarray}
 h_{00}&=&(\frac{2 M}{r}+Kr^{-2w_2})\, ,\\
 h_{ik}&=&(\frac{2 M}{r}+Kr^{-2w_2})n_i n_k\, ,\\
h_{33}&=&(\frac{2 M}{r}+Kr^{-2w_2}) \cos^2\chi 
\label{h}\, ,
\end{eqnarray}
with $\cos^2\chi=z^2/(b^2+z^2)$ and $r^2=b^2+z^2$.
Based on the above equations, we determine the first derivative of $h_{00}$ and $h_{33}$ with respect to the radial coordinate as follows:
\begin{eqnarray} 
&& \frac{dh_{00}}{dr}=-\frac {2M}{r^2}-2Kr^{-1-2w_2}w_2\,  ,\\
&& \frac{dh_{33}}{dr}=-2r^{-2(2+w_2)}(3Mr^{2w_2}+Kr(1+w_2))z^2. \ 
\end{eqnarray} 
Taking all these expressions together, we involve a plasma medium and intend to examine weak gravitational lensing in the following.

\textit{Uniform Plasma:} We first define the deflection angle of light moving around the black hole within the uniform plasma-medium distribution, which can generally be written in the following form  form~\cite{Alloqulov12025,Alloqulov2023,Alloqulov2024}
\begin{equation} \label{anglemain}
\hat{\alpha}_{uni}=\hat{\alpha}_{uni1}+\hat{\alpha}_{uni2}+\hat{\alpha}_{uni3},
\end{equation}
with
\begin{equation} \label{alphaexpansion}
\left.
\begin{aligned}
   \hat{\alpha}_{1} &= \frac{1}{2}\int_{-\infty}^{\infty} \frac{b}{r}\frac{dh_{33}}{dr} \, dz,\\ 
   \hat{\alpha}_{2} &= \frac{1}{2}\int_{-\infty}^{\infty} \frac{b}{r}\frac{1}{1-\omega^2_{e}/\omega^2}\frac{dh_{00}}{dr} \, dz,\\ 
   \hat{\alpha}_{3} &= \frac{1}{2}\int_{-\infty}^{\infty} \frac{b}{r}\Bigg(-\frac{K_{e}}{\omega^2-\omega^2_{e}}\frac{dN}{dr}\Bigg) \, dz.
\end{aligned}
\right\}
\end{equation}

Following to Eqs.~(\ref{anglemain}) and (\ref{alphaexpansion}), the deflection angle is determined by 
\begin{eqnarray}
    \hat{\alpha}_{uni}&=&\frac{1}{2}\bigg(\frac{4M}{b}+\frac{b^{-2w_{2}}K\sqrt{\pi}\Gamma[\frac{1}{2}+w_2]}{\Gamma[1+w_2]}\bigg)\nonumber\\&+&\bigg( \frac{2M}{b}+\frac{b^{-2w_{2}}K\sqrt{\pi}\Gamma[\frac{1}{2}+w_2]}{\Gamma[w_2]}\bigg)\frac{w^2}{w^2-w^2_{e}}\, . 
\end{eqnarray}

Based on the given equation, we plot the relationship between the deflection angle $\hat{\alpha}_{uni}$ and the impact parameter $b$ for various values of the parameters $w_2$, $K$, and $\omega^2_e/\omega^2$ in the spacetime of an anisotropic black hole, as shown in Fig.~\ref{fig:6}. We show how the deflection angle depends on the plasma parameters for specific values of $K$ and $w_2$ (from left to right). It is obvious that the deflection angle decreases as a function of the impact parameter $b$, while increasing the parameter $K$ results in a shift of the deflection angle towards larger values, as illustrated in Fig.~\ref{fig:6}. Interestingly, we observe that the deflection angle $\hat{\alpha}_{uni}$ is more sensitive and first decreases and increases rapidly under the impact of the parameter $\omega_2$, as shown in Fig.~\ref{fig:6}.

\textit{Non-uniform Plasma}: \textcolor{black}{In this section of our analysis, we explore the singular isothermal sphere (SIS) as the most suitable model for understanding the unique characteristics of gravitationally weak lensed photons around a black hole. Typically, an SIS is a spherical gas cloud with a central singularity where the density approaches infinity. The density distribution of an SIS is described as} \cite{Bisnovatyi-Kogan:2010flt}
\begin{equation}
    \rho(r)=\frac{\sigma^2_{\nu}}{2\pi r^2}\ ,
\end{equation}
with the one-dimensional velocity dispersion \( \sigma^2_{\nu} \), while the plasma concentration is described in the following analytical form 
\begin{equation}
    N(r)=\frac{\rho(r)}{k m_p}\, , 
\end{equation}
with the proton mass \( m_p \)  and a dimensionless constant \( k \) typically connected to dark matter. This coefficient is commonly associated with the plasma frequency as 
\begin{equation}
    \omega^2_e=K_e N(r)=\frac{K_e \sigma^2_{\nu}}{2\pi k m_p r^2}\ .
\end{equation}
Hereafter, we investigate the effect of the non-uniform plasma (SIS) on the deflection angle around an anisotropic black hole spacetime using an expression for the deflection angle given by~\cite{Al-Badawi12024,Al-Badawi2024,Alloqulov12024,Alloqulov2025,Atamurotov2022,Jiang:2024cpe}
\begin{equation}
\hat{\alpha}_{SIS}=\hat{\alpha}_{SIS1}+\hat{\alpha}_{SIS2}+\hat{\alpha}_{SIS3} \label{nonsis}\ .
\end{equation}
\textcolor{black}{Combining equations (\ref{h}), (\ref{alphaexpansion}), (\ref{nonsis})  deflection angle  can be written in the following form:
\begin{eqnarray}
   \hat{\alpha}_{SIS}&=&\frac{1}{2}\left(\frac{4M}{b}+\frac{(b^{-2w_2}K\sqrt{\pi}\Gamma[\frac{1}{2}+w_2]}{\Gamma[1+w_2]}\right)\nonumber\\&-&\frac{4 M^2 }{b \pi }\frac{w^2_{c}}{w^2}+\frac{2M}{b}+ \frac{16 M^{3} }{3 b^{3} \pi}\frac{w^2_{c}}{w^2}
\nonumber\\&+& \frac{b^{-2\omega_{2}} \, K \sqrt{\pi} \, \Gamma\!\left( \frac{1}{2} + \omega_{2} \right)}{\Gamma(\omega_{2})} 
\nonumber\\&+& \frac{4 \, b^{-2(1 + \omega_{2}}) \, K \, M^{2} \, \omega_{2} \, \Gamma\!\left( \frac{3}{2} + \omega_{2} \right) }{\sqrt{\pi} \, \Gamma\!\left( 2 + \omega_{2} \right)}\frac{w^2_{c}}{w^2}. \label{eq:alphasis}   
\end{eqnarray}
 These calculations bring up a supplementary plasma constant $\omega^2_c$ which has the following analytic expression 
 \begin{equation}
    \omega^2_c=\frac{K_e \sigma^2_{\nu}}{2\pi k m_p R^2_S}\ . 
 \end{equation}}
From Eq.~(\ref{eq:alphasis}), we examine the dependence of the deflection angle $\hat{\alpha}_{SIS}$ on the impact parameter $b$ for varying parameters $w_{2}$, $K$, $\omega^2_{c}/\omega^2$ for non-uniform plasma medium; see Fig. \ref{fig:7}. We observe a similar rate change in the behavior of the deflection angle $\hat{\alpha}_{SIS}$ for the non-uniform plasma case, similarly to what is observed in the uniform plasma case, as shown in Fig.~\ref{fig:7}. From the results, it is obvious that the magnitude of the deflection angle in the uniform plasma case is
larger than the one in the non-uniform plasma case, as clearly shown in Figs.~\ref{fig:6} and \ref{fig:7}.
\begin{figure}[!htb]
 \begin{center}
      \includegraphics[scale=0.34]{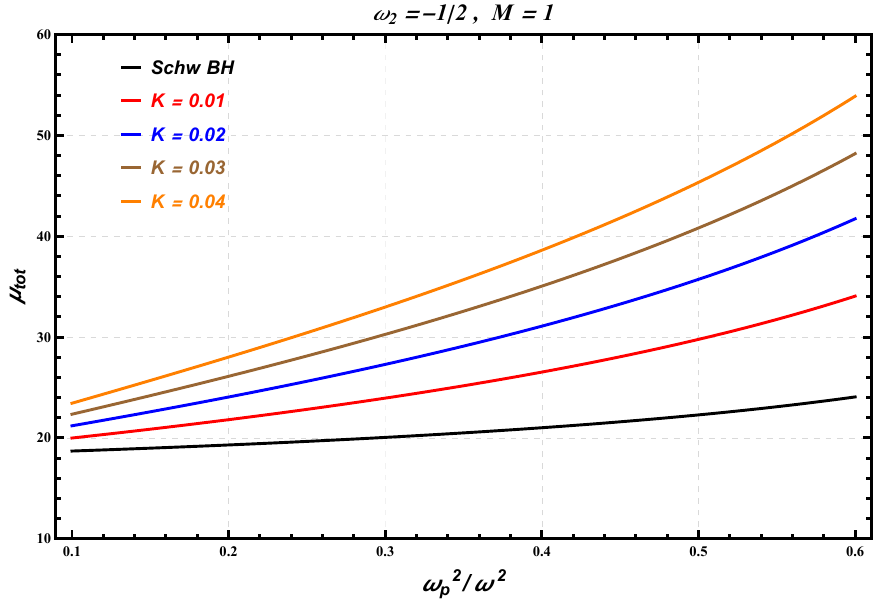}
      \includegraphics[scale=0.34]{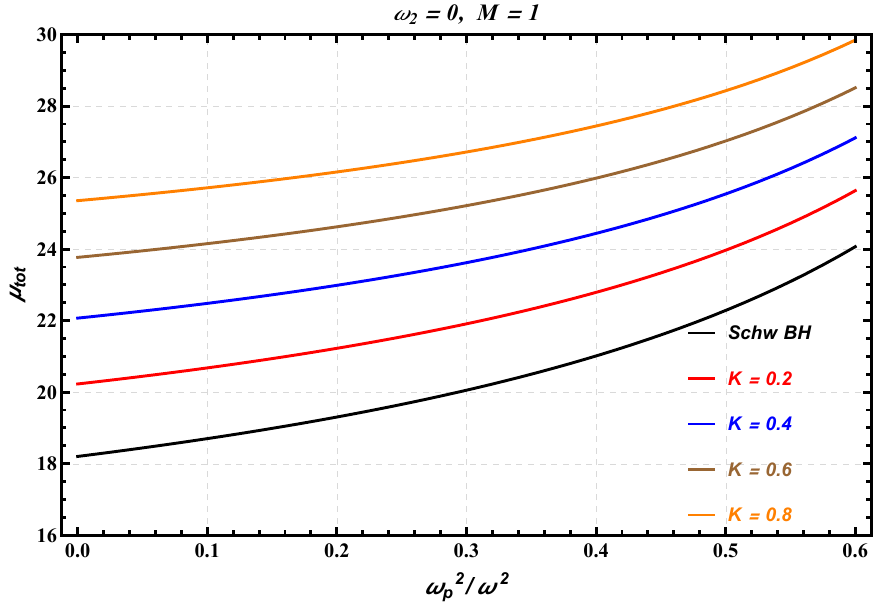}
      \includegraphics[scale=0.34]{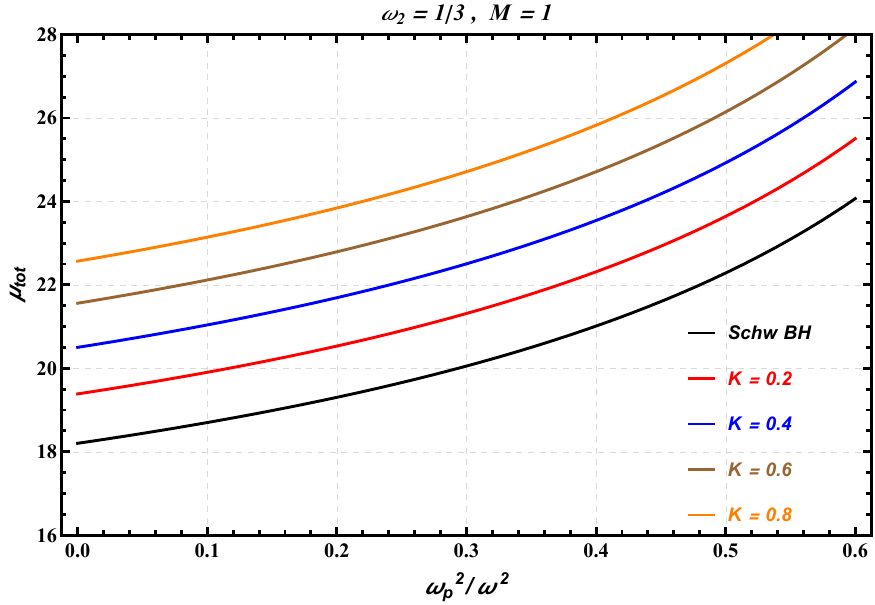}\\
   \includegraphics[scale=0.4]{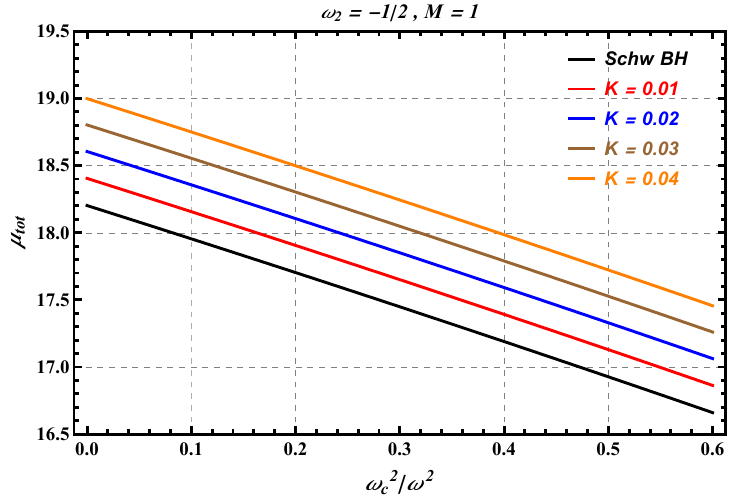}
   \includegraphics[scale=0.4]{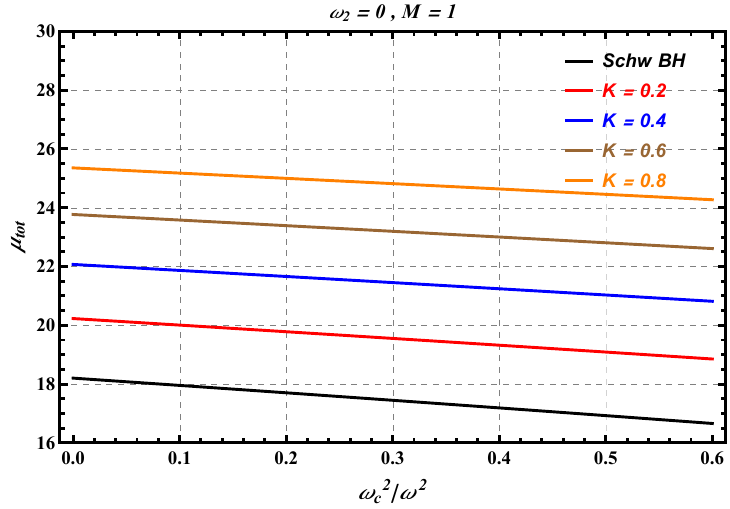}
   \includegraphics[scale=0.4]{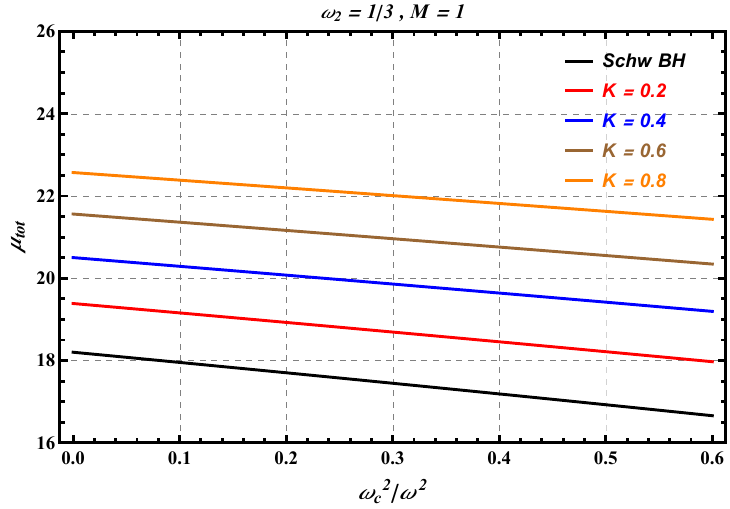}
  \end{center}
\caption{\label{fig:5} Variation of the total magnification $\mu_{\mathrm{tot}}$ with the plasma parameter for both uniform (top row) and non-uniform (bottom row) plasma distributions. The parameter $K$ is varied while keeping $w_{2}$ (from left to right) fixed and the impact parameter set to $b=3M$. 
}
\end{figure}

\section{Magnification of gravitationally lensed image}\label{Sec:magnification}

Now, we examine the brightness of the image formed in the presence of a plasma, utilizing the deflection angle of light rays around an anisotropic black hole.  By applying the lens equation, we can express the brightness in terms of the light angles around the black hole, denoted as ($\hat{\alpha}$, $\theta$ and $\beta$)\cite{Bozza2008lens}
\begin{align}\label{lenseq}
\theta D_\mathrm{s}=\beta D_\mathrm{s}+\hat{\alpha}D_\mathrm{ds}\, . 
\end{align}
Here, it is worth noting that notations $D_\mathrm{s}$, $D_\mathrm{d}$, and $D_\mathrm{ds}$ respectively represent the distances between the source and the observer, between the lens and the observer, and between the source and the lens, whereas $\theta$ and $\beta$ denote the angular positions of the image and the source. Based on the equation above, we can reformulate the expression for $\beta$ as follows:
\begin{align}\label{newlenseq}
\beta=\theta -\frac{D_\mathrm{ds}}{D_\mathrm{s}}\frac{\xi(\theta)}{D_\mathrm{d}}\frac{1}{\theta}\ ,
\end{align}
where $\xi(\theta)=|\hat{\alpha}_b|b$ and $b=D_\mathrm{d}\theta$.  
If the shape of the image looks like a ring, it is defined as Einstein's ring and radius of Einstein's ring is $R_s=D_\mathrm{d}\theta_E$. The angular $\theta_E$ due to spacetime geometry between the images of the source in a vacuum  \cite{1999grle.book.....S} can be written as
\begin{figure*}[!htb]
 \begin{center}
      \includegraphics[scale=0.355]{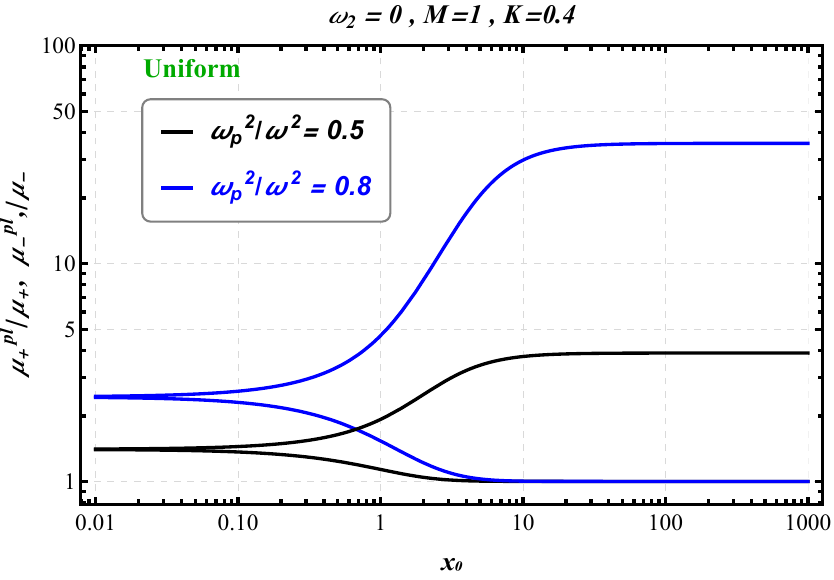}
      \includegraphics[scale=0.355]{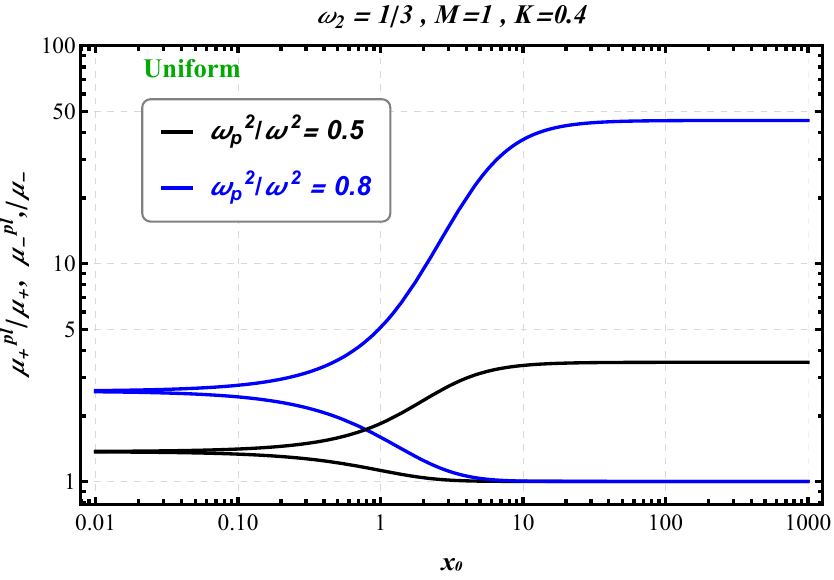}
      \includegraphics[scale=0.355]{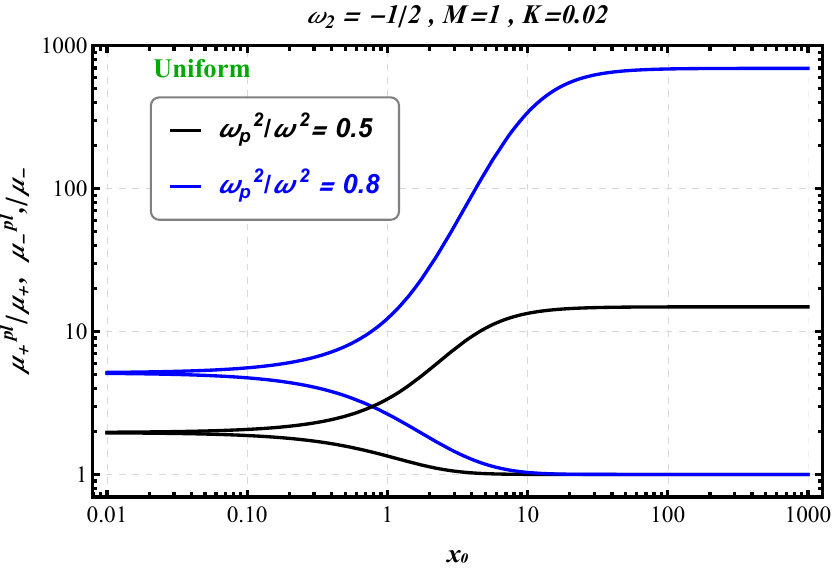}
   \includegraphics[scale=0.355]{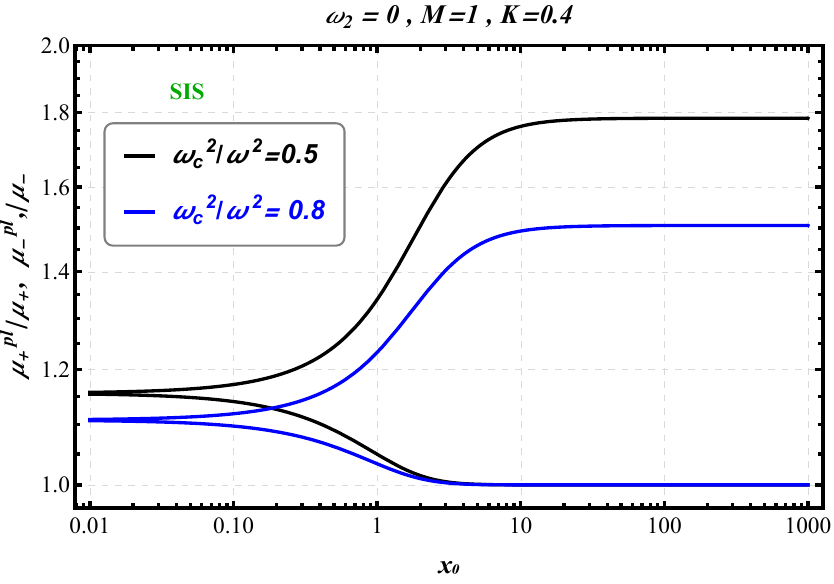}
   \includegraphics[scale=0.355]{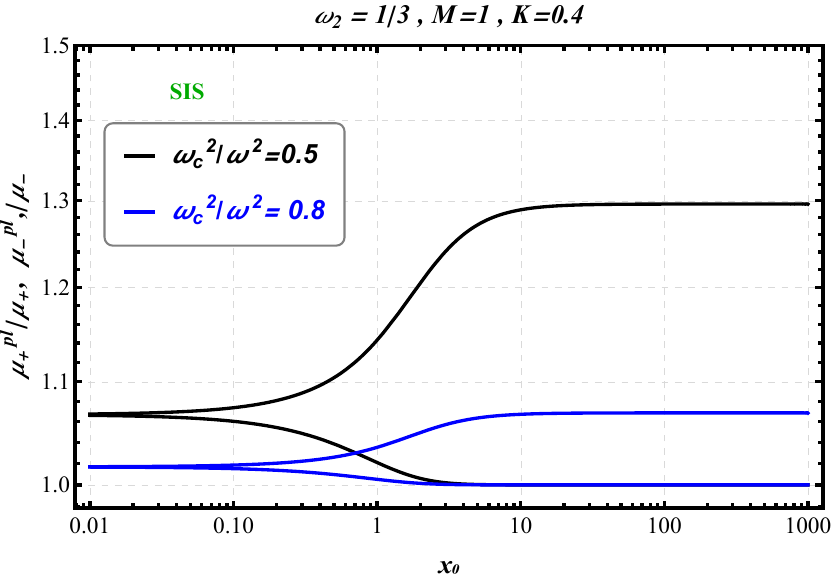}
   \includegraphics[scale=0.355]{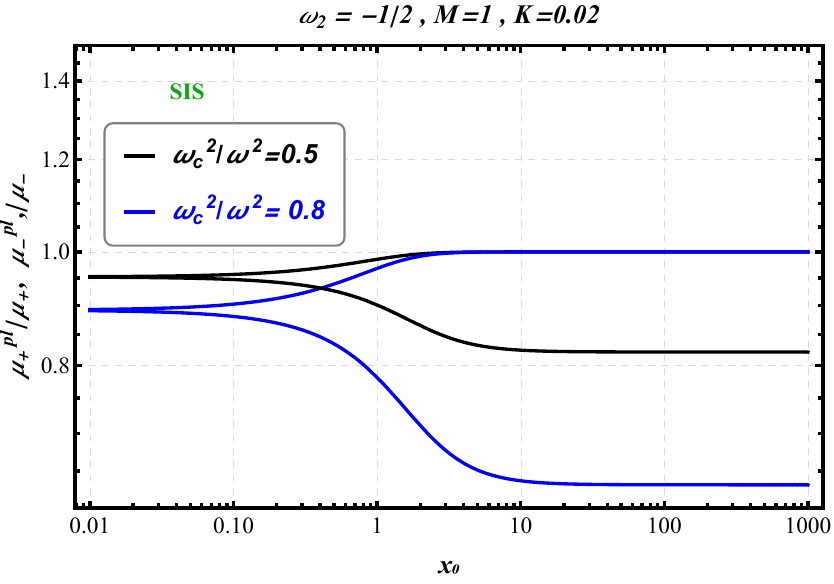}
  \end{center} 
\caption{\label{fig:9} Magnification of the lensed image as a function of the normalized angular position of the source relative to the lens $x_0$ for uniform (top row) and non-uniform SIS plasma (bottom row). Note that the impact parameter has been set $b=3M$ for $w_{2}=0$ (left), $w_{2}=1/3$ (middle) and $w_{2}=-1/2$ (right) in the case of fixed $K$. The magnification effects become stronger at larger parameter values.}
\end{figure*}
\begin{eqnarray}
\theta_E=\sqrt{2R_s\frac{D_{ds}}{D_dD_s}}\ .
\end{eqnarray}
Now, we explore the expression of the magnification of brightness 

\begin{align}\label{magni}
\mu_{\Sigma}=\frac{I_\mathrm{tot}}{I_*}=\underset{k}\sum\bigg|\bigg(\frac{\theta_k}{\beta}\bigg)\bigg(\frac{d\theta_k}{d\beta}\bigg)\bigg|, \quad k=1,2, \cdot \cdot \cdot ,  j\ ,
\end{align}

Here, $I_*$ denotes the unlensed brightness of the source, while $I_{\mathrm{tot}}$ represents the combined brightness of all images. The magnification of the source can be expressed in the following form.

\begin{align}\label{postparity}
 \mu^\mathrm{pl}_\mathrm{+}=\frac{1}{4}\bigg(\frac{x}{\sqrt{x^2+4}}+\frac{\sqrt{x^2+4}}{x}+2\bigg) ,
\end{align}

\begin{align}\label{negparity}
\mu^\mathrm{pl}_\mathrm{-}=\frac{1}{4}\bigg(\frac{x}{\sqrt{x^2+4}}+\frac{\sqrt{x^2+4}}{x}-2\bigg).
\end{align}
 {Here, $x={\beta}/{\theta_0}$ is a dimensionless parameter, and $\mu_{+}^{\mathrm{pl}}$ and $\mu_{-}^{\mathrm{pl}}$ correspond to the individual images. By applying equations (\ref{postparity}) and (\ref{negparity}), the total magnification can be expressed in the following form.}

\begin{align}\label{magtot}
\mu^\mathrm{pl}_\mathrm{tot}=\mu^\mathrm{pl}_{+}+\mu^\mathrm{pl}_{-}=\frac{x^2+2}{x\sqrt{x^2+4}}\ .
\end{align}

{We now turn to examine the magnification in an anisotropic black hole environment containing plasma, considering two types of plasma distributions, such as uniform and non-uniform.}

\textit{Uniform Plasma case}: Here, we examine how uniform plasma affects the magnification in an anisotropic black hole environment. For that we first define total magnification and angle (i.e., $\mu^{pl}_{tot}$ and $\theta^{pl}_{uni}$) as follows: 
\begin{equation}
    \mu^{pl}_{tot}=\mu^{pl}_{+}+\mu^{pl}_{-}=\dfrac{x^2_{uni}+2}{x_{uni}\sqrt{x^2_{uni}+4}} ,
\end{equation}
where 
$(\mu^{pl}_+)_{uni}$ and $(\mu^{pl}_-)_{uni}$ are determined by 
   \begin{eqnarray}
       (\mu^{pl}_+)_{uni}&=&\frac{1}{4}\left(\dfrac{x_{uni}}{\sqrt{x^2_{uni}+4}}+\dfrac{\sqrt{x^2_{uni}+4}}{x_{uni}}+2\right) ,\\
       (\mu^{pl}_-)_{uni}&=&\frac{1}{4}\left(\dfrac{x_{uni}}{\sqrt{x^2_{uni}+4}}+\dfrac{\sqrt{x^2_{uni}+4}}{x_{uni}}-2\right) .
   \end{eqnarray}
   
Taking these equations into consideration, we analyze the total magnification of the image in an anisotropic black hole environment containing plasma, $\mu^\mathrm{pl}_\mathrm{tot}$ and demonstrate it in Fig.~\ref{fig:5}. As can be observed in the bottom panels of Fig.~\ref{fig:5}, the total magnification increases as the uniform plasma ${\omega^2_{p}}/{\omega^2}$ and the parameter $K$ increase for fixed $w_{2}$ (from left to right). The total magnification as a function of $x_0$ is also demonstrated for the uniform plasma case ${\omega^2_{p}}/{\omega^2}$ for fixed values of black hole parameters, $K$ and $\omega_2$ (from left to right), as shown in the top row of Fig.~\ref{fig:9}. It is evident that, for a fixed impact parameter $b$, increasing the uniform plasma parameter results in a shift of the total magnification toward larger values. 

\textit{Non-uniform Plasma case}: Here, we study the effect of non-uniform plasma  (singular isothermal sphere (as SIS medium)) on the magnification, which can be defined by
\begin{equation}
    (\mu^{pl}_{tot})_{SIS}=(\mu^{pl}_{+})_{SIS}+(\mu^{pl}_{-})_{SIS}=\dfrac{x^2_{SIS}+2}{x_{SIS}\sqrt{x^2_{SIS}+4}}\ ,\label{eq:magsis}
\end{equation}
with
 \begin{equation}
       (\mu^{pl}_+)_{SIS}=\frac{1}{4}\left(\dfrac{x_{SIS}}{\sqrt{x^2_{SIS}+4}}+\dfrac{\sqrt{x^2_{SIS}+4}}{x_{SIS}}+2\right),\ 
\end{equation}
 \begin{equation}
       (\mu^{pl}_-)_{SIS}=\frac{1}{4}\left(\dfrac{x_{SIS}}{\sqrt{x^2_{SIS}+4}}+\dfrac{\sqrt{x^2_{SIS}+4}}{x_{SIS}}-2\right),\ 
  \end{equation}
 and 
\begin{eqnarray}
   x_{SIS}&=&\frac{\beta}{(\theta^{pl}_E)_{SIS}}\, .  
\end{eqnarray}

We then analyze the total magnification of the image in an anisotropic black hole environment containing the non-uniform plasma case, $\mu^\mathrm{pl}_\mathrm{tot}$. It is clearly seen from the bottom panels of Fig.~\ref{fig:5} that the total magnification decreases as a consequence of an increase in the non-uniform plasma parameter ${\omega^2_{p}}/{\omega^2}$, while increasing the parameter $K$ results in a slight shift of total magnification towards larger values for fixed $w_{2}$ (from left to right) and the impact parameter $b$. Subsequently, the results indicate that $\mu_{\mathrm{tot}}$ increases with the uniform plasma parameter but decreases with the non-uniform plasma parameter, with the influence of both trends becoming more significant at larger values of $K$.  Additionally, we also analyzed the total magnification as a function of $x_0$, considering the non-uniform plasma case ${\omega^2_{c}}/{\omega^2}$ for keeping black hole parameters fixed. The result is shown in the bottom row of Fig.~\ref{fig:9}. It is clearly seen that, for a fixed impact parameter b, increasing both the non-uniform plasma parameter ${\omega^2_{c}}/{\omega^2}$ and $\omega_2$ leads to a decrease in the total magnification towards smaller values compared to the uniform plasma case ${\omega^2_{p}}/{\omega^2}$. The magnification effects become stronger at larger parameter values. It should be emphasized that we also observe a comparison between the uniform and non-uniform plasma medium cases. The results clearly show that the magnification in the uniform plasma case is generally larger than in the non-uniform case, as shown in Fig.~\ref{fig:9}.

\section{Wave Dynamics in Anisotropic Black Hole Spacetimes}
\label{Sec:WaveDynamics}

\subsection{Spacetime Stability under Axial Perturbations}
\label{subsec:axial-stability}

Before analyzing wave propagation through scalar field scattering in plasma, it is important to ensure that the background spacetime itself is stable under small perturbations. A spacetime susceptible to dynamical instabilities may lead to unphysical or misleading scattering behavior. To this end, we study the response of the black hole geometry to axial (odd-parity) gravitational perturbations using the Regge–Wheeler formalism. This analysis confirms the dynamical stability of the system and justifies the use of the metric as a static background for test field propagation.

{
The stability analysis presented here has implications for multiple observational channels. While we focus on theoretical stability, the quasi-normal modes associated with stable oscillations could be probed through gravitational wave observations \cite{LIGOScientific:2016aoc} and potentially through their imprints in the interferometric signatures of black hole shadows \cite{Vincent:2022fwj}.
}

We apply Chandrasekhar’s method~\cite{Chandrasekhar:1985kt} to derive the master equation for axial perturbations, which takes the standard Regge–Wheeler form:
\begin{equation}
    \frac{d^2 \Psi}{dr_*^2} + \left[\omega^2 - V_{\text{RW}}(r) \right] \Psi = 0,
\end{equation}
where the tortoise coordinate $r_*$ is defined by
\begin{equation}
    \frac{dr_*}{dr} = \frac{1}{f(r)},
\end{equation}
and the effective potential is given by
\begin{equation}
    V_{\text{RW}}(r) = f(r) \left[ \frac{\ell(\ell + 1)}{r^2} - \frac{6M}{r^3} - \frac{K (2\omega_2)(2\omega_2 + 1)}{r^{2\omega_2 + 2}} \right].
\end{equation}

\begin{figure*}[htbp]
    \centering
    \includegraphics[width=\textwidth]{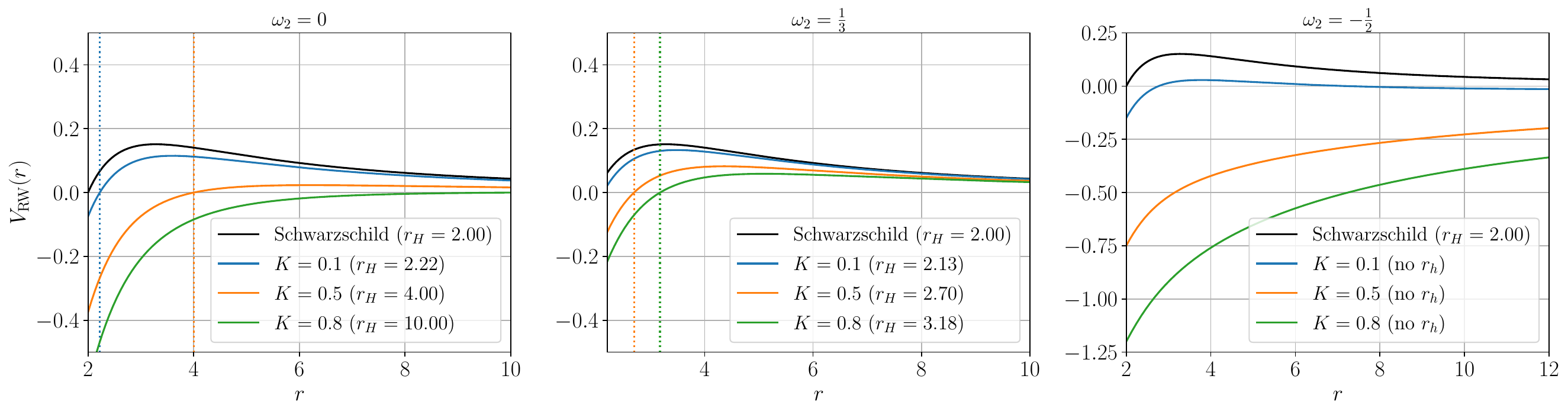}
    \caption{Regge--Wheeler effective potential \( V_{\mathrm{RW}}(r) \) for axial (odd-parity) gravitational perturbations around a black hole surrounded by anisotropic fluid. The plots are shown for three representative values of the equation-of-state parameter \( \omega_2 = 0, \frac{1}{3}, -\frac{1}{2} \). Each curve corresponds to a different value of the anisotropy parameter \( K = 0 \text{(Schwarzschild)}, 0.1, 0.5, 0.8 \). Dotted vertical lines mark the location of the event horizon \( r_H \) for each configuration. For all cases with \( \ell = 2 \), the potential remains positive outside the horizon and decays asymptotically, indicating stability under axial perturbations. The mass parameter is set to unity.
    }
    \label{Fig:regge_wheeler_potential}
\end{figure*}

As evident from \cref{Fig:regge_wheeler_potential}, the Regge–Wheeler potential remains positive outside the event horizon for all cases with \( \ell = 2 \), suggesting that the geometry is stable against axial perturbations. In particular, while the negative contribution due to the anisotropic fluid grows with $r$ for $\omega_2 = -\frac{1}{2}$, it does not overwhelm the potential barrier in physically relevant scenarios. Therefore, the black hole spacetime considered here can be treated as dynamically stable, validating our use of it as a static background in the analysis of scalar wave scattering in the next section.

\begin{figure*}[htbp]
    \centering
    \begin{minipage}[t]{0.49\textwidth}
        \centering
        \includegraphics[width=\textwidth]{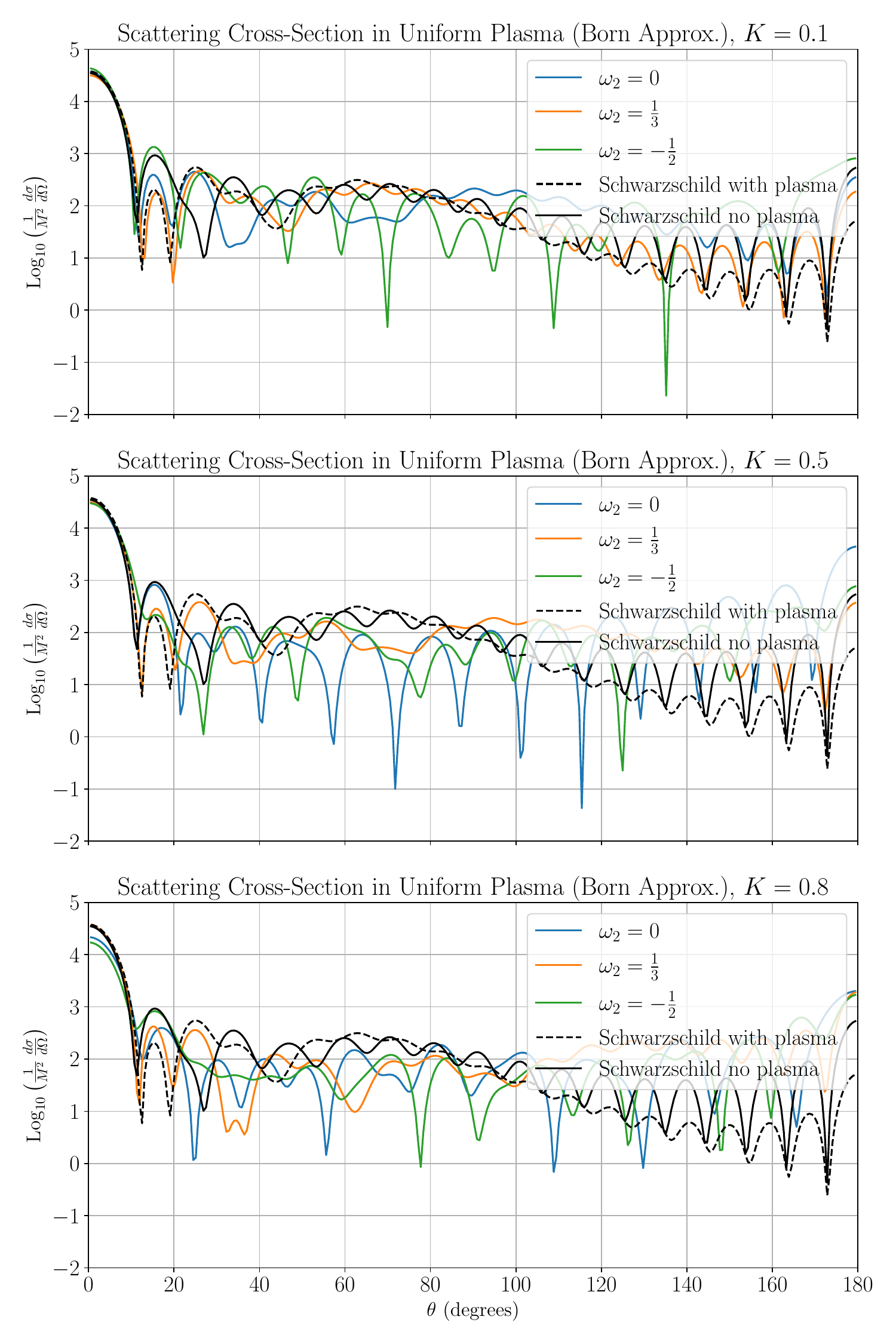}
        \\[-1ex]
        \text{(a) Uniform Plasma}
    \end{minipage}
    \hfill
    \begin{minipage}[t]{0.49\textwidth}
        \centering
        \includegraphics[width=\textwidth]{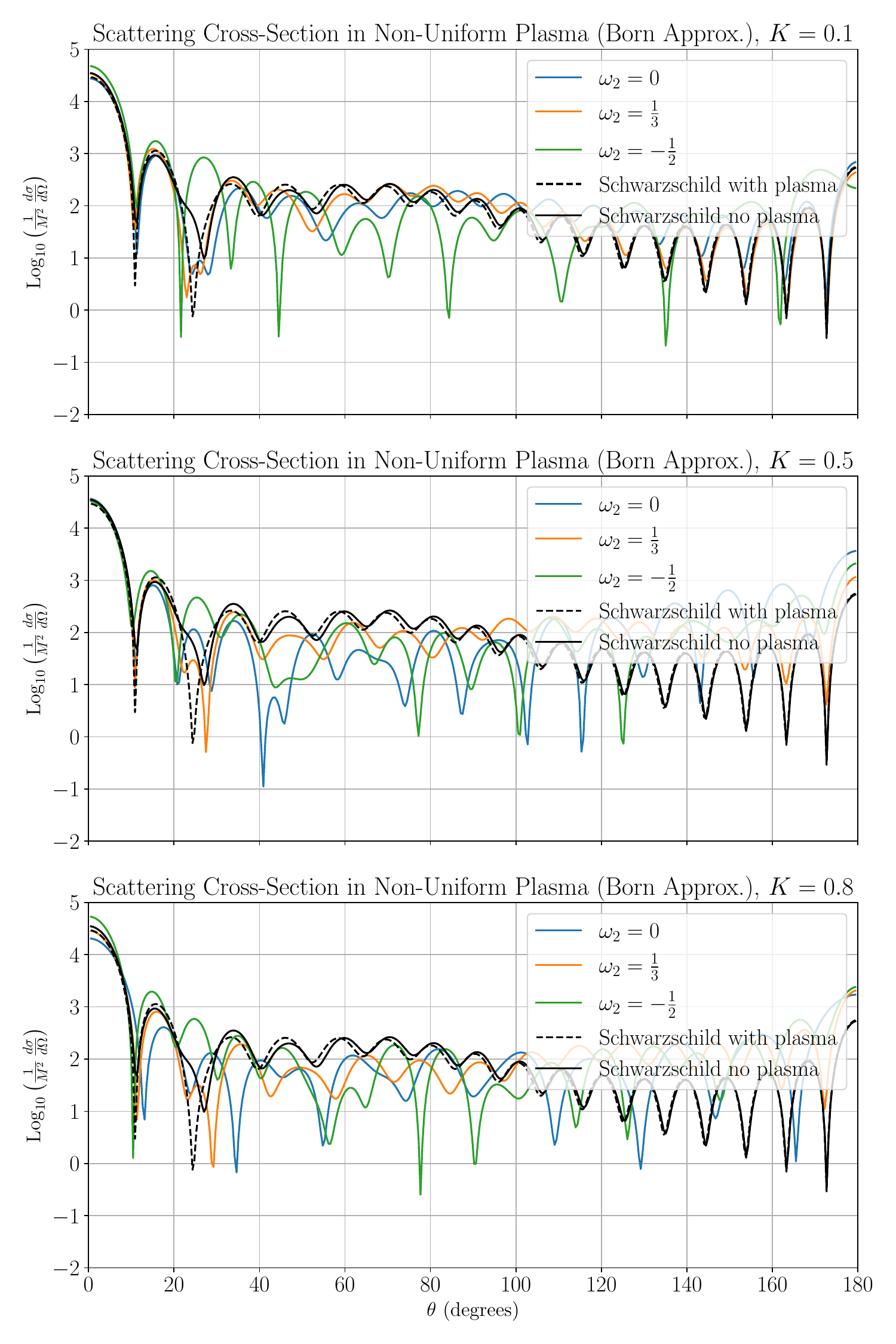}
        \\[-1ex]
        (b) Non-uniform Plasma
    \end{minipage}
    \caption{Logarithmic differential scattering cross-section \( Log_{10}\left(\frac{1}{M^2} \frac{d\sigma}{d\Omega}\right) \) as a function of scattering angle \( \theta \) (in degrees) for scalar wave scattering around an anisotropic black hole surrounded by plasma using Born approximation. The \textit{left column} corresponds to the {uniform plasma} case, while the \textit{right column} shows results for a non-uniform power-law plasma. Each row corresponds to a different value of the anisotropic fluid parameter \( K = 0.1, 0.5, 0.8 \). In each subplot, results are shown for various equation-of-state parameters \( \omega_2 \in \{0, \frac{1}{3}, -\frac{1}{2}\} \). The Schwarzschild case (\( K = 0 \)) is shown in black for reference. Phase shifts are computed using the Born approximation.}
    \label{fig:partial_scattering_uniform_vs_powerlaw_born}
\end{figure*}

\begin{figure*}[htbp]
    \centering
    \begin{minipage}[t]{0.49\textwidth}
        \centering
        \includegraphics[width=\textwidth]{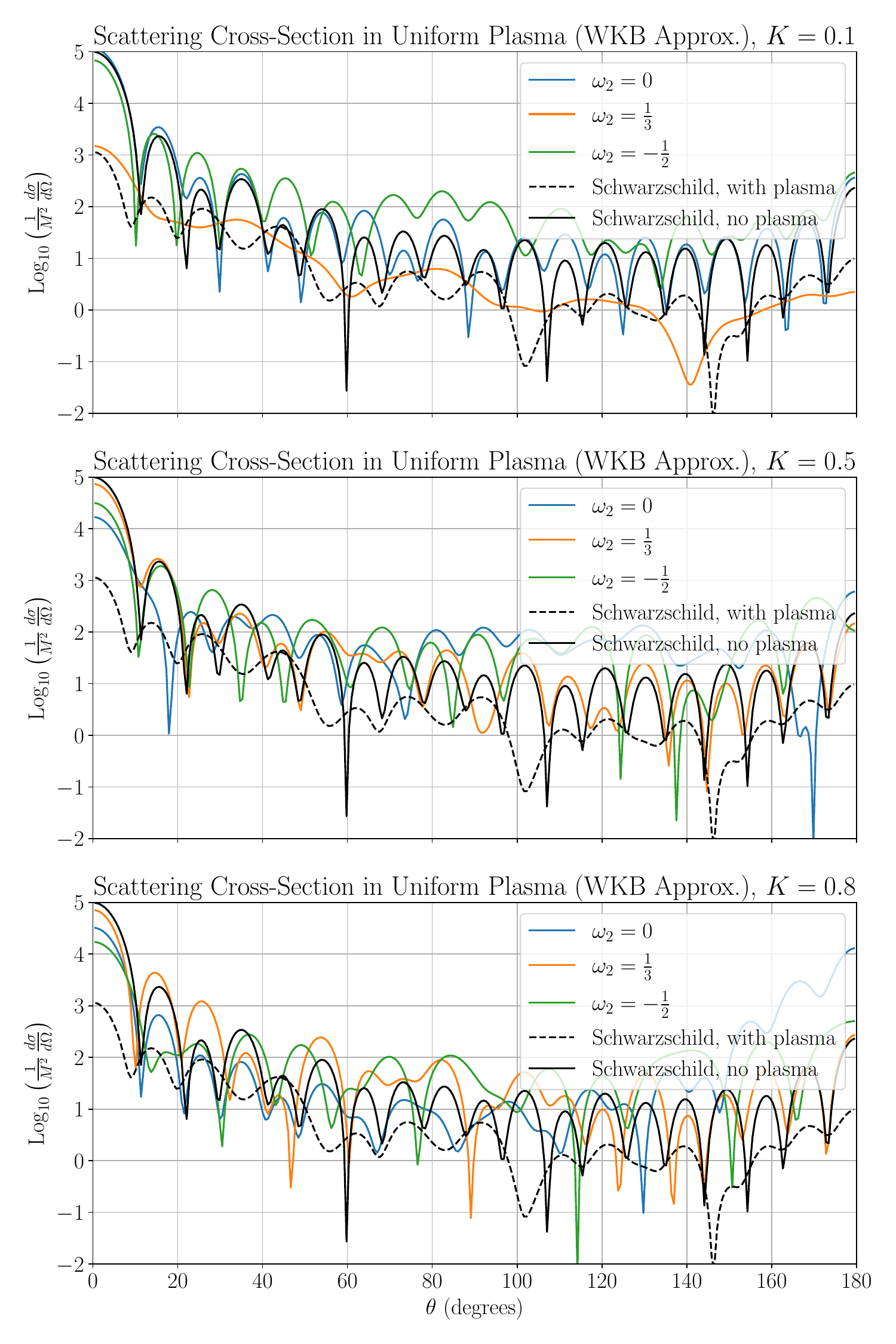}
        \\[-1ex]
        \text{(a) Uniform Plasma}
    \end{minipage}
    \hfill
    \begin{minipage}[t]{0.49\textwidth}
        \centering
        \includegraphics[width=\textwidth]{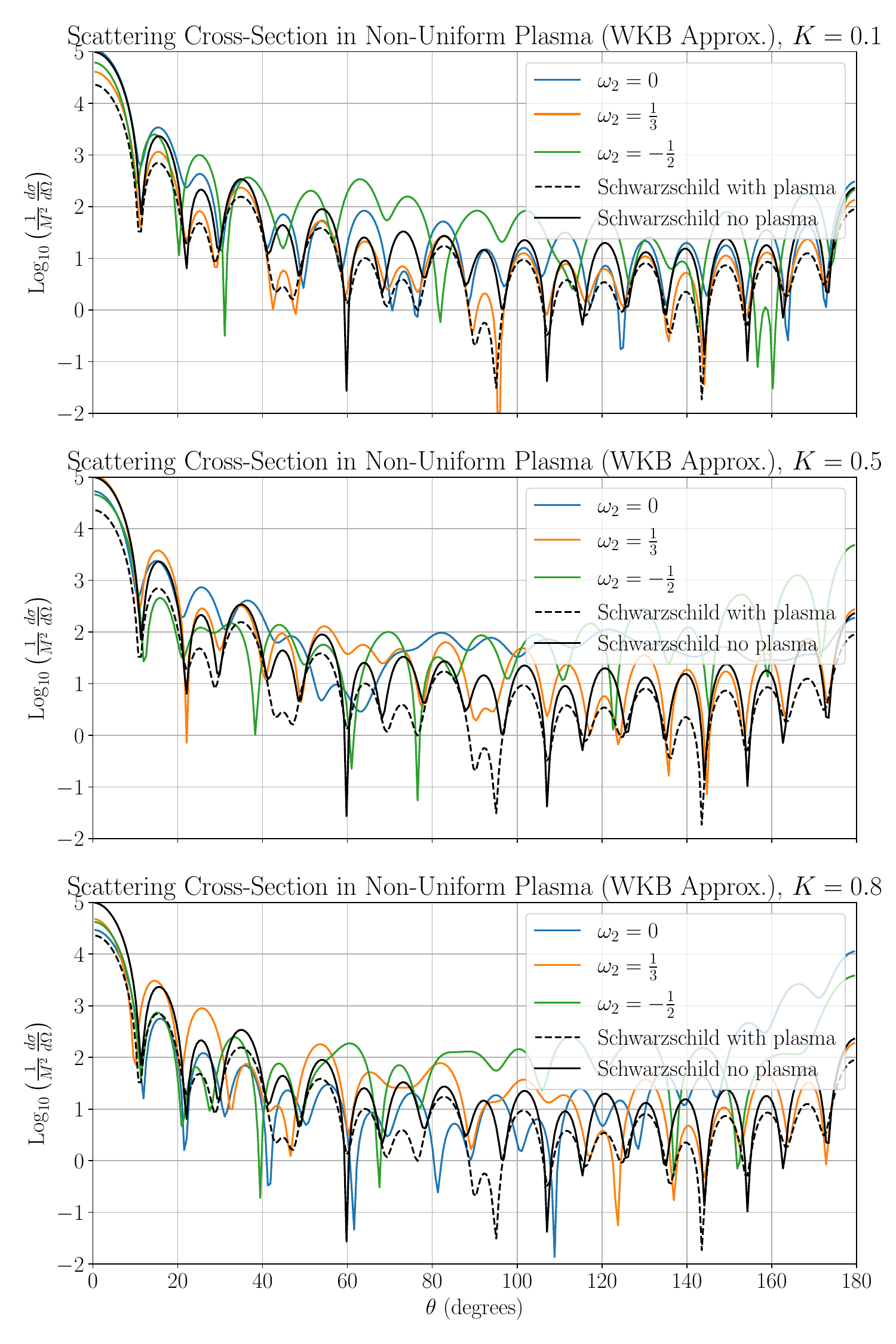}
        \\[-1ex]
        \text{(b) Non-uniform Plasma}
    \end{minipage}
    \caption{Logarithmic differential scattering cross-section \( Log_{10}\left(\frac{1}{M^2} \frac{d\sigma}{d\Omega}\right) \) as a function of scattering angle \( \theta \) (in degrees) for scalar wave scattering around an anisotropic black hole surrounded by plasma using WKB approximation. The \textit{left column} corresponds to the {uniform plasma} case, while the \textit{right column} shows results for a non-uniform power-law plasma. Each row corresponds to a different value of the anisotropic fluid parameter \( K = 0.1, 0.5, 0.8 \). In each subplot, results are shown for various equation-of-state parameters \( \omega_2 \in \{0, \frac{1}{3}, -\frac{1}{2}\} \). The Schwarzschild case (\( K = 0 \)) is shown in black for reference. Phase shifts are computed using the Born approximation.}
    \label{fig:partial_scattering_uniform_vs_powerlaw_wkb}
\end{figure*}

\subsection{Scalar Field Scattering in Plasma Surrounding the Black Hole}
\label{subsec:scalar-scattering}

We investigate the scattering of a massless scalar field in the background of a black hole surrounded by an anisotropic fluid, as introduced in \Cref{Sec:Spacetime}. Additionally, we consider the effect of a surrounding plasma medium, which modifies wave propagation through its impact on the effective refractive index. Our primary aim is to compute the differential scattering cross-section by employing partial wave analysis in both uniform and non-uniform plasma environments.

{
Our treatment of wave scattering in plasma environments builds upon established frameworks for light propagation in dispersive media \cite{Perlick:2017fio,Perlick:2023znh}. The presence of plasma introduces frequency-dependent modifications to the scattering cross-section that could be relevant for interpreting multi-wavelength observations of black hole environments.
}

\subsection{Wave Equation and Separation of Variables}

The dynamics of a massless scalar field $\Phi$ in curved spacetime is governed by the covariant Klein-Gordon equation:
\begin{equation}
    \Box \Phi = \frac{1}{\sqrt{-g}} \partial_\mu \left( \sqrt{-g}\, g^{\mu\nu} \partial_\nu \Phi \right) = 0,
\end{equation}
where $g$ is the determinant of the metric tensor $g_{\mu\nu}$. Exploiting spherical symmetry, the field can be separated into radial and angular components as:
\begin{equation}
    \Phi(t, r, \theta, \phi) = \frac{1}{r} \sum_{\ell m} \psi_\ell(r) Y_{\ell m}(\theta, \phi) e^{-i\omega t},
\end{equation}
where $Y_{\ell m}$ are the spherical harmonics and $\omega$ is the frequency as measured at spatial infinity.

\subsection{Effective Potential in the Presence of Plasma}
Let us perform the analysis of a test scalar field propagating in the same spacetime, including the effects of a surrounding plasma. This approach does not perturb the geometry but rather explores wave dynamics in a fixed curved background.

{
The modifications to the effective potential due to plasma effects follow the analytical treatments developed in recent years for handling light propagation in dispersive media around compact objects \cite{Perlick:2017fio,Feleppa:2024vdk}. These developments provide rigorous mathematical foundations for our analysis of how plasma alters wave scattering phenomena.
}

Substituting the above ansatz into the wave equation yields the radial equation:
\begin{equation}
    \frac{d^2 \psi_\ell}{dr_*^2} + \left[\omega^2 - V_\text{eff}(r) \right] \psi_\ell = 0,
\end{equation}
where the tortoise coordinate $r_*$ is defined by
\begin{equation}
    \frac{dr_*}{dr} = \frac{1}{f(r)}, \quad \text{with} \quad f(r) = 1 - \frac{2M}{r} - \frac{K}{r^{2\omega_2}}.
\end{equation}
The effective potential in the presence of plasma takes the form:
\begin{equation}\label{eq:veff}
    V_\text{eff}(r) = f(r) \left[ \frac{\ell(\ell+1)}{r^2} + \frac{f'(r)}{r} + \omega_p^2(r) \right],
\end{equation}
where $\omega_p(r)$ is the local plasma frequency that modifies the dispersion relation.

\subsection{Plasma Profiles}

To capture physically relevant scenarios, we consider two plasma models:
\begin{itemize}
    \item \textit{Uniform Plasma:} $\omega_p^2(r) = \omega_0^2$ (constant),
    \item \textit{Non-Uniform (Power-law) Plasma:} $\omega_p^2(r) = \dfrac{k}{r^h}$, where $k > 0$ and $h$ controls the falloff.
\end{itemize}
These profiles represent simplified models of plasma distributions around compact objects.

\subsection{Phase Shift Computation}

The scattering amplitude and corresponding cross-section depend critically on the phase shifts $\delta_\ell$ accumulated by each partial wave. We compute $\delta_\ell$ using two complementary methods suited for different physical regimes.

\subsubsection{Born Approximation}

In the weak potential limit, a perturbative expansion of the wave solution yields the first Born approximation:
\begin{equation}\label{eq:born}
    \delta_\ell^{\text{Born}} \approx -\frac{1}{2\omega} \int_{r_0}^{\infty} \left[ V_\text{eff}(r) - \frac{\ell(\ell+1)}{r^2} \right] \sin^2(\omega r) \, dr.
\end{equation}
This expression subtracts the centrifugal term to ensure convergence and isolates the impact of the curved background and plasma. It is most valid for small $K$, large $r$, and high-frequency waves.

\subsubsection{WKB Approximation and Amplitude-Based Phase Shift}

When the effective potential varies slowly, the semiclassical WKB method offers a more accurate alternative. The scalar wave equation can be treated as a Schrödinger-type equation with slowly varying potential. In this case, the phase shift is given by the difference in accumulated phase (amplitude) between the curved and flat space:
\begin{equation}\label{eq:wkbdelta}
    \delta_\ell^{\text{WKB}} \approx \int_{r_t}^{\infty} \left[ \sqrt{\omega^2 - V_\text{eff}(r)} - \sqrt{\omega^2 - \frac{\ell(\ell+1)}{r^2}} \right] dr,
\end{equation}
where $r_t$ is the classical turning point defined by $V_\text{eff}(r_t) = \omega^2$. The integrand measures the shift in wavenumber due to curvature and plasma effects. This formulation directly captures the amplitude difference accumulated by the wave and is particularly effective for large $\ell$ and smooth potentials.

In regions where $\omega^2 < V_\text{eff}(r)$, the integrand is set to zero to reflect classically forbidden regions.

\subsection{Differential Scattering Cross-Section}

The angular distribution of scattered waves is given by the partial wave expansion:
\begin{equation}\label{eq:dcs}
    \frac{d\sigma}{d\Omega} = \left| f(\theta) \right|^2 = \left| \sum_{\ell=0}^{\ell_\text{max}} \frac{2\ell+1}{2i\omega} \left( e^{2i\delta_\ell} - 1 \right) P_\ell(\cos\theta) \right|^2,
\end{equation}
where $P_\ell$ are Legendre polynomials. The sum is truncated at a finite $\ell_\text{max}$ depending on the energy $\omega$ and required precision. To facilitate comparison across scales, we define a normalized cross-section:
\begin{equation}
    \text{Log}_{10} \left( \frac{1}{M^2} \frac{d\sigma}{d\Omega} \right).
\end{equation}

\subsection{Numerical Implementation and Results}

We numerically evaluate the phase shifts using both Eqs.~\eqref{eq:born} and \eqref{eq:wkbdelta} for various choices of $K$, $\omega_2$, and plasma profile. The cross-sections are then computed via Eq.~\eqref{eq:dcs} and plotted as a function of scattering angle. \cref{fig:partial_scattering_uniform_vs_powerlaw_born,fig:partial_scattering_uniform_vs_powerlaw_wkb} illustrates these results for increasing $K$, comparing anisotropic matter cases to the Schwarzschild limit ($K=0$).

\subsection{Comparison of Approximation Methods}

To understand the reliability and domain of validity for both methods, we compare their computational and theoretical characteristics (see table \ref{tab:Born_WkB_data}). This comparison highlights how each approximation performs across different physical regimes, especially in the presence
of anisotropic matter and plasma.

\begin{table}[htbp]
    \centering
    \renewcommand{\arraystretch}{1.2}
    \footnotesize
    \begin{tabular}{|>{\centering\arraybackslash}p{2.2cm}|>{\centering\arraybackslash}p{2.9cm}|>{\centering\arraybackslash}p{2.9cm}|}
        \hline
        \textbf{Aspect} & \textbf{Born Approximation} & \textbf{WKB Approximation} \\
        \hline
        Regime of Validity & Weak potential, far-field & High-frequency, smooth potential \\
        \hline
        Accuracy at low $\ell$ & Poor & Moderate \\
        \hline
        Accuracy at high $\ell$ & Moderate & Excellent \\
        \hline
        Plasma Effects & Perturbative & Included via turning point \\
        \hline
        Computational Cost & Low & Moderate to High \\
        \hline
        Horizon/strong field & Invalid & Partially valid \\
        \hline
    \end{tabular}
    \caption{Comparison between Born and WKB approximations.}
    \label{tab:Born_WkB_data}
\end{table}

As shown in table \ref{tab:Born_WkB_data}, the Born approximation is computationally efficient and provides
qualitative insight into the scattering features at high frequencies and large radii, but it
systematically underestimates contributions from strong curvature regions and cannot reliably
capture low-$\ell$ behavior. In contrast, the WKB approximation incorporates curvature and
plasma effects more accurately through the turning-point structure of the potential. This
makes it well suited for high multipole orders and regimes where the potential varies slowly,
although at the cost of increased computational complexity


\subsection{Physical Interpretation and Scattering Features}

Our analysis of the differential scattering cross-sections in \cref{fig:partial_scattering_uniform_vs_powerlaw_born,fig:partial_scattering_uniform_vs_powerlaw_wkb} shows that anisotropic matter alters the effective potential and thereby modifies the scattering profile, with concrete trends governed by the anisotropic fluid parameter \(K\), the equation-of-state parameter \(\omega_2\), and the surrounding plasma:
\begin{itemize}
    \item For $\omega_2 = 0$, the scattering profile remains closest to the Schwarzschild case. Increasing $K$ suppresses oscillatory amplitudes, particularly at larger angles.
    \item For $\omega_2 = \tfrac{1}{3}$ (radiation-like fluid), forward scattering is consistently enhanced. At higher $K$, oscillations become smoother in the Born approximation but remain well resolved in the WKB approximation.
    \item For $\omega_2 = -\tfrac{1}{2}$, a strong enhancement in backward scattering is evident. This effect grows with increasing $K$, producing sharper peaks in the WKB spectra around $\theta \sim 120^\circ - 180^\circ$.
\end{itemize}
The impact of plasma is also evident:
\begin{itemize}
    \item \textit{Uniform plasma} reduces the contrast of oscillations, damping interference features across all $K$. The suppression is most visible at small angles, where forward scattering dominates.
    \item \textit{Non-uniform plasma} introduces pronounced modulations and shifts in the oscillatory pattern. These effects are especially strong for $\omega_2 = -\tfrac{1}{2}$, where additional resonant-like peaks appear in the large-angle region.
\end{itemize}
A comparison of approximation schemes highlights clear differences:
\begin{itemize}
    \item \textit{Born Approximation:} Captures the overall angular decay but consistently underestimates the height of interference peaks and the magnitude of backscattering. This is most notable for $\omega_2 = -\tfrac{1}{2}$.
    \item \textit{WKB Approximation:} Provides a more accurate description of interference fringes, especially at large angles. For non-uniform plasma, the WKB spectra exhibit well-resolved resonant oscillations absent in the Born approximation.
\end{itemize}
The Schwarzschild case ($K=0$), with and without plasma, serves as the reference baseline. Deviations from this baseline confirm that both anisotropy and plasma effects significantly reshape scattering signatures: anisotropy chiefly governs angular redistribution (forward vs. backward scattering), while plasma dictates the sharpness and modulation of interference structures.

{
The scattering features we identify could potentially be constrained through complementary observational approaches. X-ray reflection spectroscopy \cite{Bambi:2016sac,Bambi:2020jpe} provides constraints on the inner accretion flow geometry, while techniques based on the orbital angular momentum of light \cite{Tamburini:2021lyi,Tamburini:2021jok} offer alternative probes of spacetime structure. The frequency-dependent scattering signatures we compute may be relevant for interpreting multi-wavelength observations of black hole environments.
}

\begin{figure*}[ht]
    \centering
    \includegraphics[width=0.95\textwidth]{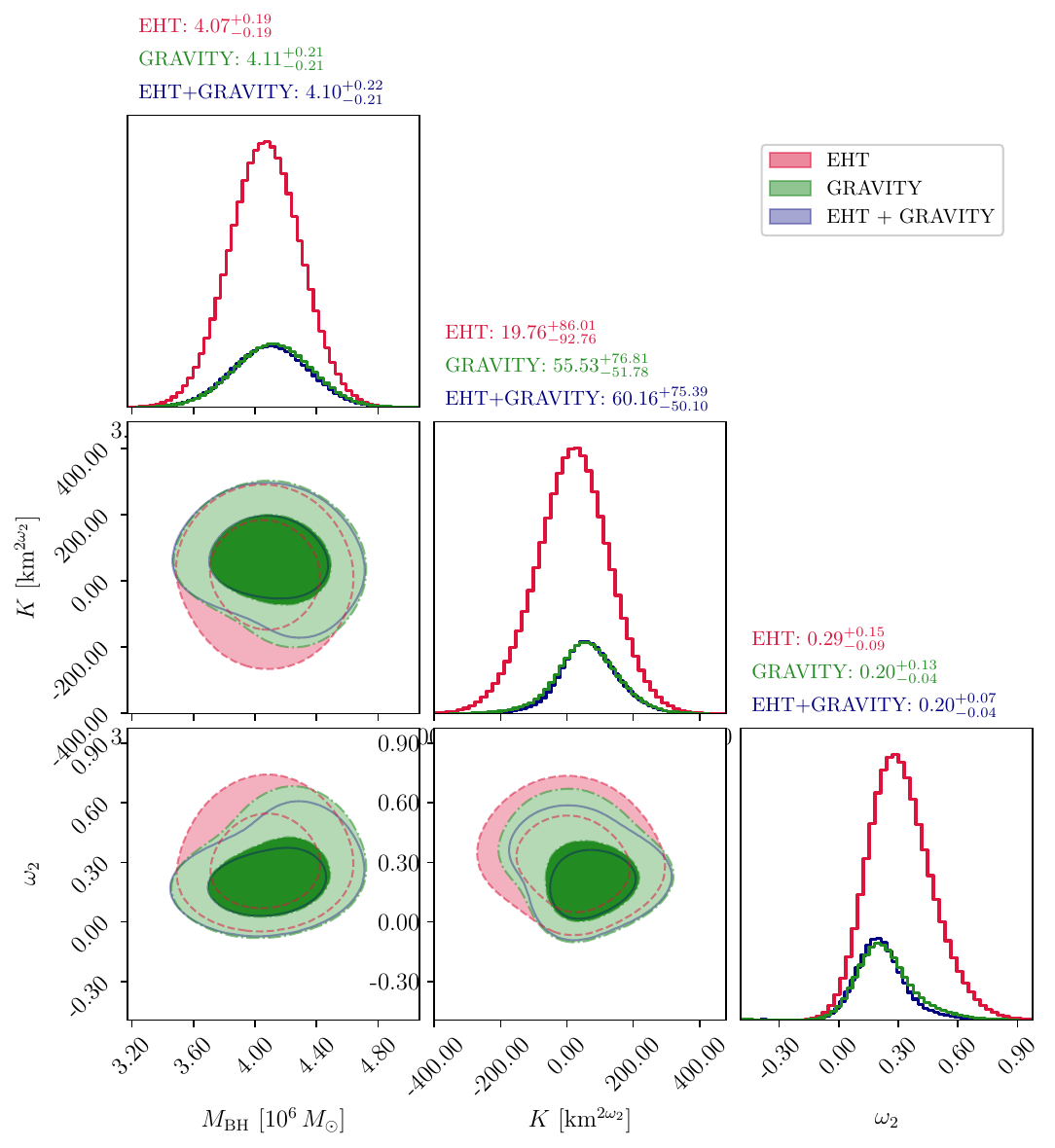}
    \caption{Corner plot showing marginalized posterior distributions and correlations among the black hole mass \( M \), anisotropy parameter \( \omega_2 \), and fluid strength parameter \( K \) for the Sgr A* black hole. Constraints are shown separately for EHT, GRAVITY, and the combined EHT+GRAVITY datasets.}
    \label{fig:corner_sgra}
\end{figure*}

\begin{figure*}[ht]
    \centering
    \includegraphics[width=0.95\textwidth]{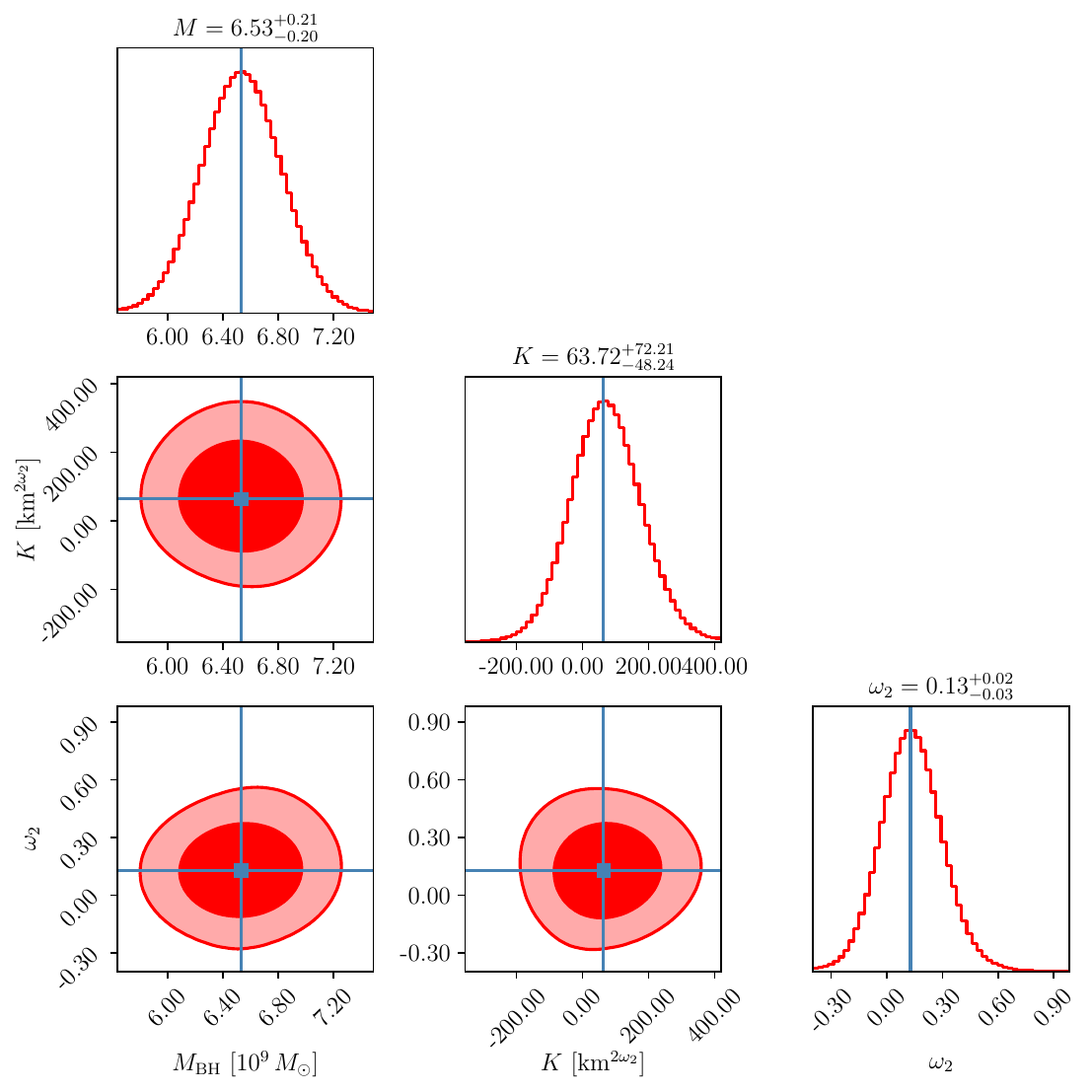}
    \caption{Corner plot showing marginalized posterior distributions and parameter correlations for the M87* black hole within the anisotropic fluid spacetime model. The plot includes constraints on the black hole mass \( M \), anisotropy parameter \( \omega_2 \), and the fluid strength parameter \( K \), inferred using EHT observational data.}
    \label{fig:corner_m87}
\end{figure*}

\section{Parameter Estimation for Anisotropic Fluid Black Hole Spacetime}
\label{sec:param_estimation_anisotropic}

To constrain the mass \( M \), fluid parameter \( K \), and equation of state parameter \( \omega_2 \) of the anisotropic fluid black hole (AFBH), we employed a Bayesian parameter estimation approach using Markov Chain Monte Carlo (MCMC) methods implemented via the \texttt{emcee} package \cite{Foreman-Mackey_2013}. This analysis was performed using three distinct observational datasets for SgrA*: (1) Event Horizon Telescope (EHT), (2) GRAVITY collaboration data, and (3) a combined dataset (EHT + GRAVITY). For M87*, due to limited GRAVITY data, we used only the EHT observations.

The observable quantity used for parameter inference is the angular diameter of the shadow \( \theta_{\text{sh}} \), which is related to the theoretical photon sphere radius and the metric function \( f(r) \) as:
\begin{equation}
    R_{\text{sh}} = r_{\text{ph}} \sqrt{\frac{1}{f(r_{\text{ph}})}} \, ,
\end{equation}
where the photon sphere radius \( r_{\text{ph}} \) satisfies:
\begin{equation}
    \left. \frac{d}{dr} \left( \frac{r^2}{f(r)} \right) \right|_{r = r_{\text{ph}}} = 0 \, .
\end{equation}
This condition ensures unstable circular photon orbits.

The predicted angular shadow size is given by:
\begin{equation}
    \theta_{\text{sh}} = R_{\text{sh}} \times \frac{2M \times 4.847 \times 10^{-6}}{D} \, ,
\end{equation}
where \( D \) is the distance to the black hole (16.8 Mpc for M87*, and 0.0083 Mpc for SgrA*).

The log-likelihood function used in MCMC sampling is:
\begin{equation}
    \log \mathcal{L}(M, K, \omega_2) = -\frac{1}{2} \sum_i \left( \frac{\theta_{\text{sh}}^{\text{theory}}(M, K, \omega_2) - \theta_{\text{sh}}^{\text{obs}}}{\sigma_i} \right)^2 \, ,
\end{equation}
where \( \theta_{\text{sh}}^{\text{obs}} \) and \( \sigma_i \) denote the observed value and its uncertainty, respectively.
With observed shadow sizes and associated errors taken from the respective datasets.

We adopted informative Gaussian priors for all parameters:
\begin{equation}
\begin{aligned}
    & M \sim \mathcal{N}(4.01 \times 10^6, (0.2 \times 10^6)^2) \\
    & K \sim \mathcal{N}(0, 100^2) \\
    & \omega_2 \sim \mathcal{N}(0.2, 0.2^2)
\end{aligned}
\end{equation}

The posterior was obtained via Bayes’ theorem:
\begin{equation}
    P(M, K, \omega_2 \mid \mathcal{D}) \propto \mathcal{L}(\mathcal{D} \mid M, K, \omega_2) \cdot \pi(M, K, \omega_2) \, .
\end{equation}

We used 16 walkers and ran 50,000 steps, discarding the first 1,000 as burn-in and thinning by a factor of 30. The posterior distributions for SgrA* are shown in Fig.~\ref{fig:corner_sgra}, comparing the EHT-only, GRAVITY-only, and combined results.

\subsection{Estimated Parameters}

The median and 68\% confidence intervals obtained from the MCMC analysis are:

\paragraph{SgrA* (EHT-only)}
\begin{align}
    M &= 4.07^{+0.19}_{-0.19} \times 10^6 M_\odot \, , \\
    K &= 19.76^{+86.01}_{-92.76} \, , \\
    \omega_2 &= 0.29^{+0.15}_{-0.09} \, .
\end{align}

\paragraph{SgrA* (GRAVITY-only)}
\begin{align}
    M &= 4.11^{+0.21}_{-0.21} \times 10^6 M_\odot \, , \\
    K &= 55.53^{+76.81}_{-51.78} \, , \\
    \omega_2 &= 0.20^{+0.13}_{-0.04} \, .
\end{align}

\paragraph{SgrA* (EHT + GRAVITY Combined)}
\begin{align}
    M &= 4.10^{+0.22}_{-0.21} \times 10^6 M_\odot \, , \\
    K &= 60.16^{+75.39}_{-50.10} \, , \\
    \omega_2 &= 0.20^{+0.07}_{0.04} \, .
\end{align}
For M87*, we only used the EHT data due to limited GRAVITY information. The best-fit values obtained are:
\paragraph{M87* (EHT-only)}
\begin{align}
    M &= 6.53^{+0.21}_{-0.20} \times 10^9 M_\odot \, , \\
    K &= 63.72^{+74.21}_{-48.24} \, , \\
    \omega_2 &= 0.13^{+0.02}_{-0.03} \, .
\end{align}

These results illustrate the strength of combining multi-instrument observations to improve parameter constraints. Notably, the addition of GRAVITY data tightens the credible intervals and helps reduce degeneracy, particularly in the \(M\)-\(K\) plane. The constraints for M87* are also in good agreement with prior EHT-based estimates, reinforcing the viability of the anisotropic fluid model as a descriptor of supermassive black hole spacetimes.

\subsection{Discussion}

Our results demonstrate how combining different types of observations can significantly improve constraints on black hole parameters.
The EHT \cite{EventHorizonTelescope:2019dse,EventHorizonTelescope:2022wkp} data alone imposes strong bounds on the shadow radius, which is directly sensitive to the spacetime geometry near the photon sphere. In contrast, the GRAVITY \cite{GRAVITY:2021xju,GRAVITY:2024tth} dataset constrains the black hole mass by tracking the orbital motion of surrounding stars and hot spots, probing the geometry at larger radial distances. The combination of these two data sources reduces degeneracy in the parameter space, particularly tightening the joint confidence regions in the \( (K, \omega_2) \) plane.

{
Our parameter estimation results should be interpreted in the context of what is actually measurable with current VLBI arrays. The constraints we derive on anisotropic fluid parameters through lensing observables would manifest as specific modifications to visibility-domain signatures \cite{Vincent:2022fwj,Aratore:2021usi,Feleppa:2025ejh}. Future high-resolution observations with next-generation VLBI instruments could provide more stringent constraints on these parameters through detailed analysis of interferometric data.
}

For Sgr~A*, the joint analysis (EHT + GRAVITY) improves the precision of all parameters. However, the parameter \( K \), which quantifies the strength of anisotropic fluid corrections to the Schwarzschild potential, remains weakly constrained for two reasons. First, its effects become significant only at intermediate radii—between the ISCO and photon sphere—where current observations lack sufficient resolution. Second, a functional degeneracy in the correction term \( K / r^{2\omega_2} \) allows different parameter combinations to reproduce similar signatures. This is reflected in the broad posteriors and the tilt of confidence contours. Moreover, the positive best-fit value of \( \omega_2 \) suggests that the effective anisotropic fluid around Sgr~A* behaves more like a radiation-like medium, supporting its physical consistency with known forms of energy in strong gravity regimes.

For M87*, where only EHT data is available, our constraints are consistent with the EHT Collaboration’s reported mass estimate, \( (6.5 \pm 0.7) \times 10^9 M_\odot \). The inferred nonzero value of \( K \) points to possible deviations from standard Schwarzschild geometry, while the positive best-fit value of \( \omega_2 \approx 0.13 \) likewise suggests that the effective fluid surrounding the black hole resembles a radiation-dominated medium rather than dark energy. This interpretation aligns with a scenario in which near-horizon matter content modifies the vacuum geometry while remaining compatible with current observational bounds.

These findings support the viability of anisotropic fluid models in describing deviations from standard vacuum solutions. Further improvements in observational precision—such as those expected from the next-generation EHT (ngEHT) and GRAVITY+ upgrades—will be essential to break parameter degeneracies and probe the nature of matter in strong gravity regimes. Together, these results highlight both the promise and the present limitations of anisotropic fluid models, setting the stage for our concluding remarks.

\section{Conclusion}\label{sec:conlusion}

In this work, we have systematically analyzed the physical and observational consequences of black hole spacetimes surrounded by an anisotropic fluid within the framework of general relativity. The main findings can be summarized as follows:

\begin{enumerate}

    \item \emph{Spacetime structure:}  
    \begin{itemize}
        \item The black hole spacetime is described by a modified Schwarzschild-like metric with an additional term governed by two parameters: \( K \), quantifying the anisotropy strength, and \( \omega_2 \), describing the radial fall-off of the fluid. These parameters encode deviations from vacuum solutions and allow modeling of environments ranging from radiation- to dark-energy-like distributions.
        \item Horizon structure analysis shows that the anisotropic matter satisfies the null and weak energy conditions across a wide parameter space (Figs.~\ref{fig:anisotropic_metric_K_scan}, \ref{fig:anisotropic_BH_shaded}).
    \end{itemize}

    \item \emph{Photon trajectories and shadows:}  
    \begin{itemize}
        \item Photon trajectories were investigated in the presence of uniform and non-uniform (power-law) plasma profiles. Equations of motion were derived and photon and black hole shadow radii were computed.
        \item Photon sphere radii increase with higher \( K \) for dust (\(\omega_2=0\)), radiation (\(\omega_2=1/3\)), and dark energy-like (\(\omega_2=-1/2\)) cases (Fig.~\ref{fig:ph}).
        \item Plasma environments significantly modify the photon sphere, with larger radii at higher densities. The black hole shadow radii also increase with \( K \) (Fig.~\ref{fig:shadow}).
        \item Observational constraints on \( K \) in the plasma-medium environment were examined using data from M87* and Sgr A* (Fig.~\ref{fig:constraint}).
    \end{itemize}

    \item \emph{Gravitational lensing:}  
    \begin{itemize}
        \item The anisotropic fluid parameter \( K \) enhances the deflection angle \(\hat{\alpha}\), while \(\omega_2\) further modifies its radial dependence, producing faster decreases in deflection (Figs.~\ref{fig:6}, \ref{fig:7}).
        \item Plasma medium has a strong effect: deflection is larger in uniform plasma than in non-uniform (SIS) plasma, indicating high sensitivity of gravitational lensing to the plasma environment.
        \item Lensing magnification was studied and shown to increase with larger \( K \) and \(\omega_2\) (Figs.~\ref{fig:5}, \ref{fig:9}).
    \end{itemize}

    \item \emph{Wave dynamics and stability:}  
    \begin{itemize}
        \item Dynamical stability of the spacetime was assessed using axial (odd-parity) gravitational perturbations via Chandrasekhar’s method. The Regge–Wheeler potential remains positive for all physically relevant \( K \) and \(\omega_2\), indicating linear stability (Fig.~\ref{Fig:regge_wheeler_potential}).
        \item Scalar wave scattering was analyzed using Born and WKB approximations. Anisotropic parameters and plasma profiles imprint distinct interference patterns on scattered waves (Figs.~\ref{fig:partial_scattering_uniform_vs_powerlaw_born}, \ref{fig:partial_scattering_uniform_vs_powerlaw_wkb}).
    \end{itemize}

    \item \emph{Observational constraints and parameter estimation:}  
    \begin{itemize}
        \item Using a Bayesian MCMC framework, black hole mass and anisotropic fluid parameters were constrained:
        \begin{enumerate}
        \item \emph{Sgr~A*}: Joint EHT and GRAVITY analysis yields  
            \( M = 4.10^{+0.22}_{-0.21} \times 10^6 M_\odot \),  
            \( K = 60.16^{+75.39}_{-50.10} \),  
            \( \omega_2 = 0.20^{+0.07}_{-0.04} \) (Fig.~\ref{fig:corner_sgra}).  
            The joint analysis reduces degeneracies in \( (K, \omega_2) \) and supports a radiation-like effective fluid near the black hole.
            
            \item \emph{M87*}: EHT-only analysis yields  
            \( M = 6.53^{+0.21}_{-0.20} \times 10^9 M_\odot \),  
            \( K = 63.72^{+74.21}_{-48.24} \),  
            \( \omega_2 = 0.13^{+0.02}_{-0.03} \) (Fig.~\ref{fig:corner_m87}).  
            These results are consistent with previous mass estimates and indicate small deviations from Schwarzschild geometry. Positive \(\omega_2\) values support a radiation-like medium rather than a dark-energy-like component.
        \end{enumerate}
    \end{itemize}

\end{enumerate}

In the future, observations from the advanced EHT and GRAVITY collaborations are expected to significantly enhance our understanding of black hole environments. The combination of EHT’s shadow measurements and GRAVITY’s precise dynamical mass estimates is particularly powerful for Sgr~A*, reducing degeneracies in the inferred anisotropy parameters. Upcoming facilities such as the next-generation EHT (ngEHT) and the Extremely Large Telescope (ELT) will provide higher-resolution near-horizon observations, potentially revealing subtle signatures of exotic matter distributions and further testing anisotropic fluid models.

Overall, this study presents a unified framework linking anisotropic fluid modifications of black hole spacetimes to multiple observational channels—including gravitational lensing, wave scattering, and parameter estimation. The results demonstrate that anisotropies and plasma leave detectable imprints on astrophysical observables, providing a solid theoretical basis for testing alternative matter configurations near black holes with current and upcoming instruments.

\section*{Acknowledgements}
PS expresses gratitude to the Vellore Institute of Technology for financial support through its Seed Grant (No. SG20230079, Year 2023). Additionally, PS and HN acknowledge the support of the Anusandhan National Research Foundation (ANRF) through the Science and Engineering Research Board (SERB) Core Research Grant (Grant No. CRG/2023/008980).

\bibliographystyle{JHEP}
\bibliography{references}
\end{document}